\pdfoutput=1

\documentclass{svjour3}                    
\usepackage{setspace}   
\usepackage{lipsum} 
\usepackage{appendix}
\usepackage{amsmath}
\usepackage{graphicx}
\usepackage{lineno}
\usepackage{array}
\usepackage{longtable}
\usepackage{xcolor}
\usepackage{rotating}
\usepackage{multirow}
\usepackage{multicol}
\usepackage{a4wide} 
\usepackage[printwatermark]{xwatermark}
\usepackage{subcaption}
\usepackage{wrapfig}

\makeatletter
\renewcommand\paragraph{\@startsection{paragraph}{4}{\z@}
            {-2.5ex\@plus -1ex \@minus -.25ex}
            {0ex \@plus .25ex}
            {\normalfont\normalsize\emph}}
\makeatother
\newcommand{\mybox}[1]{
\vspace{3mm}
\noindent 
\framebox[\linewidth][c]{
\parbox[b]{0.95\linewidth}{
{#1}
}
}
}
\usepackage{marginnote}

\usepackage[backgroundcolor=white,bordercolor=blue,linecolor=blue]{todonotes}

\journalname{}
\date{}
\newcommand*\patchAmsMathEnvironmentForLineno[1]{
\expandafter\let\csname old#1\expandafter\endcsname\csname #1\endcsname
\expandafter\let\csname oldend#1\expandafter\endcsname\csname end#1\endcsname
\renewenvironment{#1}
{\linenomath\csname old#1\endcsname}
{\csname oldend#1\endcsname\endlinenomath}}
\newcommand*\patchBothAmsMathEnvironmentsForLineno[1]{
\patchAmsMathEnvironmentForLineno{#1}
\patchAmsMathEnvironmentForLineno{#1*}}
\AtBeginDocument{
\patchBothAmsMathEnvironmentsForLineno{equation}
\patchBothAmsMathEnvironmentsForLineno{align}
\patchBothAmsMathEnvironmentsForLineno{flalign}
\patchBothAmsMathEnvironmentsForLineno{alignat}
\patchBothAmsMathEnvironmentsForLineno{gather}
\patchBothAmsMathEnvironmentsForLineno{multline}
}

\begin{document}

\title{Test Case Selection and Prioritization Using Machine Learning: A~Systematic Literature Review}

\author{Rongqi Pan     \and
        Mojtaba Bagherzadeh \and 
        Taher A. Ghaleb \and 
        Lionel Briand
}

\institute{Rongqi Pan, Mojtaba Bagherzadeh, Taher A. Ghaleb, and Lionel Briand\at Nanda Lab, School of Electrical Engineering and Computer Science (EECS), University of Ottawa, Canada\\
Emails: \{rpan099,m.bagherzadeh,tghaleb,lbriand\}@uottawa.ca
}

\maketitle

\begin{abstract}
Regression testing is an essential activity to assure that software code changes do not adversely affect existing functionalities. With the wide adoption of Continuous Integration (CI) in software projects, which increases the frequency of running software builds, running all tests can be time-consuming and resource-intensive. To alleviate that problem, Test case Selection and Prioritization (TSP) techniques have been proposed to improve regression testing by selecting and prioritizing test cases in order to provide early feedback to developers. In recent years, researchers have relied on Machine Learning (ML) techniques to achieve effective TSP (ML-based TSP). Such techniques help combine information about test cases, from partial and imperfect sources, into accurate prediction models. This work conducts a systematic literature review focused on ML-based TSP techniques, aiming to perform an in-depth analysis of the state of the art, thus gaining insights regarding future avenues of research. To that end, we analyze 29 primary studies published from 2006 to 2020, which have been identified through a systematic and documented process. This paper addresses five research questions addressing variations in ML-based TSP techniques and feature sets for training and testing ML models, alternative metrics used for evaluating the techniques, the performance of techniques, and the reproducibility of the published studies. We summarize the results related to our research questions in a high-level summary that can be used as a taxonomy for classifying future TSP studies.

\end{abstract}

\section{Introduction}

Regression testing is an essential software quality assurance activity to gain confidence that changes in software code have not adversely affected existing functionalities.
Nowadays, many software projects adopt Continuous Integration (CI), a practice that runs software builds, including regression testing, automatically and more frequently~\cite{fowler2006continuous}.
By default, for a software system with a small codebase, all previously executed test cases are run (run-them-all approach) when they are still applicable (not obsolete). However, due to the frequent execution of CI builds and the prominence of large codebases, in practice, the run-them-all approach can be time-consuming and resource-intensive, requiring many servers and hours or even days to complete~\cite{khatibsyarbini2018test,lima2020test,ghaleb2019empirical}. This limitation warrants the need for practical techniques that seek to reduce the effort required for regression testing in various ways, including Test case Selection and Prioritization (TSP), which is our main focus in this paper.

TSP techniques deal with the costly execution time of regression testing by selecting and prioritizing test cases that are (1) sufficient to test new changes while accounting for their side effects and (2) able to detect faults as early as possible. These techniques often rely on multiple information sources including coverage information analysis (e.g.,~\cite{rothermel2001prioritizing}), heuristics based on test execution history (e.g.,~\cite{kim2002history}), and domain-specific heuristics and rules (e.g.,~\cite{rothermel2001prioritizing}). In order to combine these factors, in recent years, researchers have increasingly relied on Machine Learning (ML) techniques to drive Test case Selection and Prioritization (ML-based TSP). Combining features of test cases from different sources can lead to improved accuracy for test case selection and prioritization. The adoption of CI helps generate richer datasets about test cases that can be used to feed and train ML models.

Lima et~al.~\cite{lima2020test} reported an increasing trend in the number of TSP techniques that explore the use of ML to combine different data sources for building appropriate feature sets.
While there are several surveys and Systematic Literature Reviews (SLR) that study and classify TSP techniques, they do not provide a detailed account and analysis of ML-based TSP. To remedy this issue, as there are now a sizable number of primary studies we can rely on, we conduct an SLR that exclusively focuses on ML-based TSP techniques, aiming for a more detailed analysis of their usage and performance. Inspired by Khatibsyarbini et~al.~\cite{khatibsyarbini2018test} and Lima et~al.~\cite{lima2020test}, we follow a typical four-step SLR process~\cite{kitchenham2004procedures,kitchenham2009systematic}. These steps are (1) the definition of research questions, (2) a search strategy (including the selection of literature repositories and search strings), (3) inclusion and exclusion criteria, and (4) a data synthesis and extraction procedure. The search resulted in 29 relevant, primary studies, which are then analyzed to address five research questions summarized below. We also provide a high-level summary of ML-based TSP, which can be used as a taxonomy for classifying future TSP studies.

\begin{itemize}
    \item The main ML techniques used for TSP are: supervised learning (ranking models), unsupervised learning (clustering), reinforcement learning, and natural language processing (RQ1).
    
    \item ML-based TSP techniques mainly rely on features that are easy to compute and based on data that are practical to collect in a CI context, including execution history, coverage information, code complexity, and textual data (RQ2).
    
    \item ML-based TSP techniques are evaluated using a variety of metrics that are, sometimes, calculated differently in TS and TP, making it difficult to compare their results. Most of the currently available subjects have extremely low failure rates, making them unsuitable for evaluating ML-based TSP techniques (RQ3). 

    \item Comparing the performance of ML-based TSP techniques is challenging due to the variation of evaluation metrics, test suite sizes, and failure rates across studies. Reporting failure rates alongside performance values helps provide more interpretable results to the wider research community (RQ4).
    \item Only six out of the 29 selected studies (21\%) can be considered reproducible, thus raising methodological issues in the studies and a lack of confidence in reported results (RQ5). 
\end{itemize}

The rest of this paper is organized as follows. Section~\ref{sec:reserachmethod} presents the research method. Section~\ref{sec:result} reports and analyzes the results for each research question. Section~\ref{sec:discussion} discusses the implications of our findings. Sections~\ref{sec:relatedwork} and \ref{sec:threats} describe the related work and threats to validity, respectively. Section~\ref{sec:conlcusion} concludes our systematic literature review.

\section{Research Method}
\label{sec:reserachmethod}
\begin{figure}[ht]
    \centering
    \includegraphics[width=0.95\columnwidth]{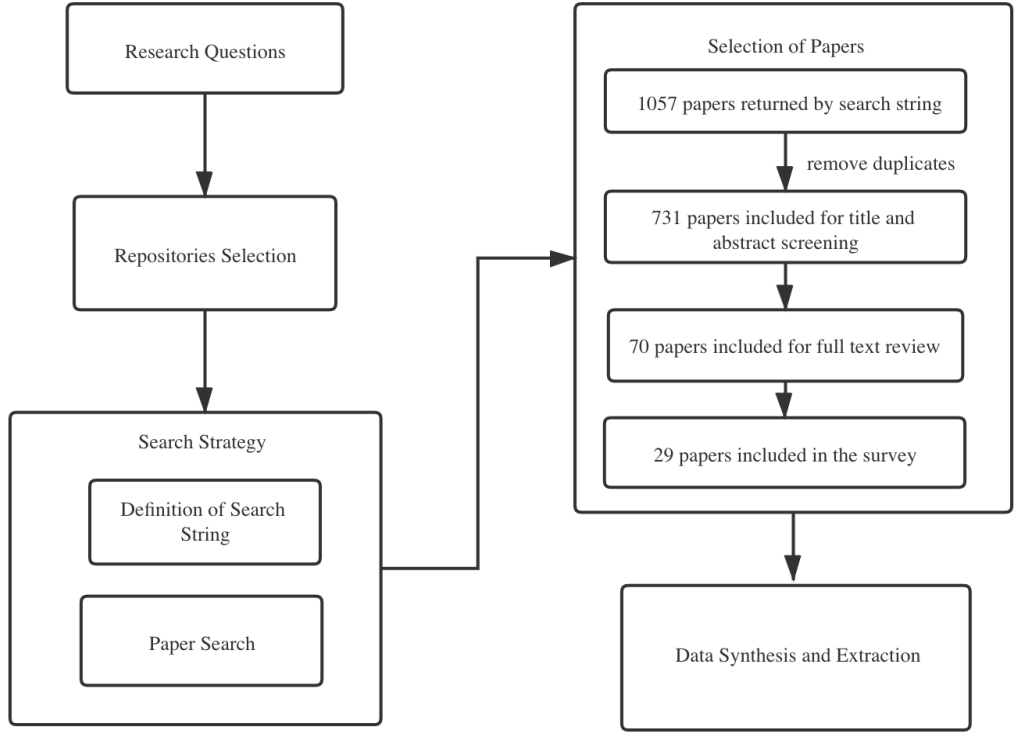}
    \caption{The Main Steps of the SLR}
    \label{fig:steps1}
\end{figure}

We conduct a Systematic Literature Review (SLR) on the application of Machine Learning (ML) techniques to Test case Selection and Prioritization (TSP). We aim to (a) analyze how ML techniques have been used, (b) assess the results they have achieved, and (c) study their limitations. In this section, we discuss the steps of the research method we carried out, which is inspired by Khatibsyarbini et~al.~\cite{khatibsyarbini2018test} and Lima et~al.~\cite{lima2020test}. Figure~\ref{fig:steps1} depicts the steps of our SLR process. The steps include the definition of research questions, search strategy (including the selection of literature repositories and search strings), study selection based on inclusion and exclusion criteria, and data synthesis and extraction. A summary of the search results is also included.

\subsection{Definition of Research Questions}
\label{rqs_definition}
To analyze and gain insights into the application of ML techniques for TSP, we address the following five research questions.

\begin{itemize}
    \item [\textbf{RQ1.}] What ML techniques are used for TSP and what are the reported strengths and weaknesses of using these techniques? The aim of this question is to identify all the ML techniques used for TSP and classify them according to categories reflecting the type of techniques. This question also aims to discuss the strengths and limitations of each ML technique as reported in the studies.

    \item [\textbf{RQ2.}] What are the features used by ML-based TSP techniques? This question focuses on identifying the types of features used by ML-based TSP techniques. Features, in turn, determine the complexity and cost of data collection. We collect the name and definition of each feature set, along with information regarding its data collection mechanism.

    \item [\textbf{RQ3.}] How are ML-based TSP techniques evaluated? This question focuses on the empirical methodology followed for evaluating the performance of ML-based TSP techniques. We collect information about the evaluation metrics, the subjects used for the experiments and their availability, and general experimental design aspects.

    \item [\textbf{RQ4.}] What is the performance of ML-based TSP techniques? This question focuses on assessing what can be currently achieved with respect to ML-based TSP. We collect the reported results, based on empirical evidence, along with the test suite size, feature sets, and ML techniques of each study.

    \item [\textbf{RQ5.}] Are ML-based TSP studies repeatable and reproducible? We investigate whether identical experiments can be rerun and whether reported results and conclusions can be validated. More specifically, we identify whether the results are provided in sufficient detail, the availability of the datasets and relevant scripts (e.g., scripts for testing and training ML models), and whether the experimental design, configurations, and procedures are presented in a sufficiently precise way.
\end{itemize}

\subsection{Search Strategy}
In this section, we explain our search strategy, including the selection of literature repositories and search strings. 

\subsubsection{Literature Repository Selection}

We performed our search in digital libraries and we chose online repositories based on the popularity and the degree of relevance to software engineering. The chosen repositories are listed in Table~\ref{tab:searchresult}.

These repositories include well-known conferences, workshops, symposiums, and journal articles. They also provide many leading software engineering publications and they are extensively used by previous survey papers in software engineering. For example, Khatibsyarbini et~al.~\cite{khatibsyarbini2018test} included all the online repositories we used for their survey of test case prioritization approaches in regression testing. Lima et~al.~\cite{lima2020test} included the ACM digital library, Scopus, Science Direct, IEEE Xplore, and Wiley online library to the repositories in their survey of Test case Prioritization in CI environments.

\begin{table}[hbt!]
\centering
\caption{Search Result for Each Repository}
\begin{tabular}{ p{3.5cm}p{2.5cm} }
 \hline
 Source &  Found papers by search query\\
 \hline
     Scopus               & 403 \\
     IEEE Xplore          & 149\\
     Ei Compendex         & 146\\
     MIT Libraries        & 100\\
     Web of Science       & 91\\
     ACM Digital Library  & 77\\
     EBSCOhost            & 68\\
     Wiley online library & 15\\
     Science Direct       & 8\\
 \hline
\end{tabular}
\label{tab:searchresult}
\vspace{-15pt}
\end{table}

\subsubsection{Search Strings}

We formulated search strings based on the goals of this work and the research questions. Our aim was to find papers that apply ML techniques for the purpose of TSP. Thus, we included general terms about the scope of this paper, such as `test', `selection', `prioritization', and `rank' in the search string. To include all ML techniques used for TSP, we also included specific ML-related terms, such as `learning', `clustering', `logistic', `support vector machine', `neural network', `bayes', `natural language processing' `k nearest neighbors', and `reinforcement learning' in the search string.
Given that the number of search terms is limited in some online repositories, we used a partial set of the terms in one search and then the remaining set in another search.
TSP techniques can also be applied for other types of testing (e.g., unit testing), but our focus is regression testing. Thus, to narrow the search, we added the term `regression' to the search string.

We noticed that some repositories require an adaptation of search strings because of differences in search engines among repositories. For example, for the MIT libraries, when we included `regression' AND (`learning' OR `clustering' OR `logistic' OR `support vector machine' OR `neural network' OR `bayes' OR `natural language processing' OR `k nearest neighbors') in the \textit{All Text} field, and included `test' AND (`selection' OR `prioritization' OR `rank') in the \textit{Title} field, we obtained too many results (678 results). Therefore, 
we only used `regression' AND (`learning' OR `clustering' OR `logistic' OR `support vector machine' OR `neural network' OR `bayes' OR `natural language processing' OR `k nearest neighbors') in the \textit{Abstract} field and included `test' AND (`selection' OR `prioritization' OR `rank') in the \textit{Title}. As a result, we obtained 100 results. For Scopus, in order to narrow the search, we limited the subjects of the results to `computer science', `engineering' and `mathematics'. We also limited the publication stage of the results to `final' to include only papers that have already been assigned to a publication volume and issue.

Table~\ref{tab:searchresult} shows the search results for each source. The highest number of papers (403) was found in Scopus. In total, 1,057 papers were found from the online repositories.

\subsection{Inclusion and Exclusion Criteria}
Table~\ref{tab:inclusion} presents the inclusion criteria we used for paper selection. In summary, we reviewed studies that are long papers ($>$ six pages), written in English, available online, in a final publication stage, and related to regression testing. We verified each study that fits the goal of our study (ML-based TSP). We excluded the papers that conduct a survey, SLR, or systematic mapping. However, we discuss them as related work (Section~\ref{sec:relatedwork}). Further, we did not set the publication period for each source but rather included all the papers in the search results. 

\begin{table}[ht]
\centering
\caption{Inclusion Criteria}
\begin{tabular}{ll}
 \hline
 &Inclusion Criteria \\
 \hline
 1 & The paper is in English\\
 2 & The paper is related to regression testing\\
 3 & The paper is related to test case selection or test case prioritization\\
 4 & The paper uses ML techniques\\
 5 & The paper is a long paper ($>$ six pages)\\
 6 & The paper is in a final publication stage\\
 7 & The paper is not a survey, a systematic literature review, or a systematic mapping\\
 \hline
\end{tabular}
\label{tab:inclusion}
\vspace{-15pt}
\end{table}

\subsection{Data Synthesis and Extraction Method}
In this section, we discuss the information regarding search results and extraction methods. 

We used Covidence tool\footnote{\url{https://www.covidence.org} to screen the obtained papers. 1,057 papers were imported for screening and, after removing duplicate articles, 731 papers were included after the title and abstract screening step. After excluding the papers based on the inclusion and exclusion criteria, 70 papers remained for a full-text manual review step.} Finally, 29 papers were included in our study. 
We then extracted related information from the papers according to our research questions. Table~\ref{tab:rqs_data} presents the data collected for each research question. The type of data extracted reflects the goal that each research question addresses.
We provide a replication package~\cite{replicationpackage} of our SLR containing the search strings and the data extracted from papers.

\begin{table}
\centering
\caption{Data Collection for Each Research Question}
\begin{tabular}{ll}
 \hline
 Research Question & Type of Data Extracted\\
 \hline
 RQ1 &  ML techniques used for TSP and their reported strengths and limitations\\
 RQ2 &  Feature sets used for prediction and data collection procedures\\
 RQ3 &  Evaluation metrics, subjects, and experiment designs\\
 RQ4 &  Performance results (accuracy) of ML-based TSP techniques\\
 RQ5 &  Availability of datasets, code, scripts, and detailed results\\
 \hline
\end{tabular}
\label{tab:rqs_data}
\end{table}

\begin{figure}
    \centering
    \includegraphics[width=0.6\columnwidth]{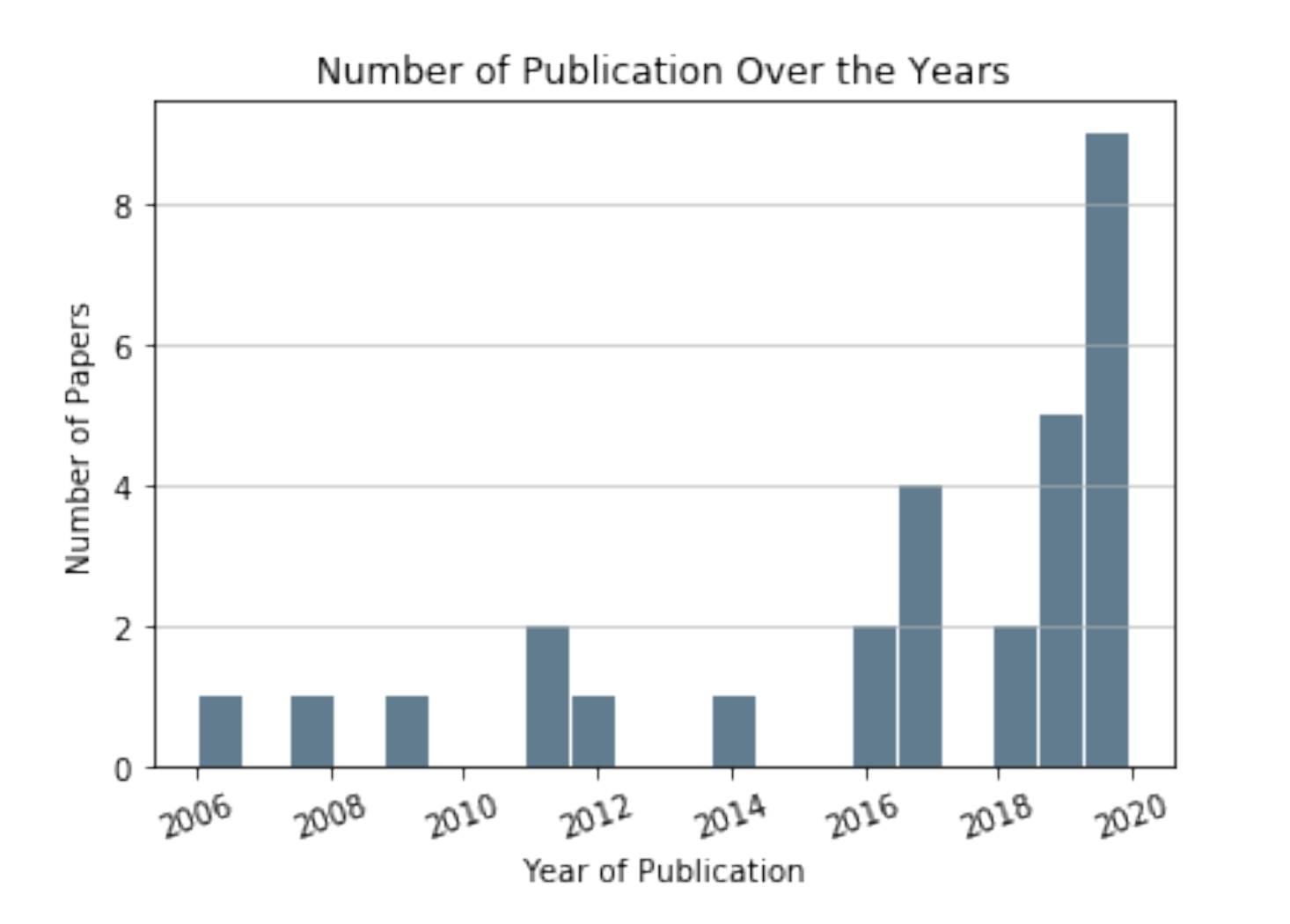}
    \caption{Number of Publications Over the Years}
    \label{fig:publication_year}
\end{figure}

There were 16 conference papers, nine journal articles, and four symposium papers. More specifically, we retrieved two papers from the IEEE International Conference on Software Maintenance (ICSM), two papers from the International Conference on Predictive Models and Data Analytics in Software Engineering (PROMISE), two papers from the ACM SIGSOFT International Symposium on Foundations of Software Engineering (FSE), and one paper from the International Conference on Software Engineering (ICSE).
We also retrieved papers from the IEEE Transactions on Software Engineering (TSE) and the Empirical Software Engineering Journal (EMSE).
They are all well-known and credible publication venues related to software engineering.
Figure~\ref{fig:publication_year} shows the number of publications over the years. We observe a positive increment from 2006 to 2020, with a rapid increase after 2018. The trend suggests that ML techniques for TSP have been a popular trend in recent years.

\section{Results}
\label{sec:result}
In this section, we address each of the five research questions defined in Section~\ref{rqs_definition} regarding the application of ML techniques for test case selection and prioritization.

\subsection{\textbf{RQ1.} What ML techniques are used for TSP and what are the reported strengths and weaknesses of using these techniques?}

As shown in Table~\ref{tab:techniques}, based on the applied ML techniques, we classify the selected papers into four groups: Supervised Learning (SL)~\cite{james2013introduction}, Unsupervised Learning (UL)~\cite{james2013introduction}, Reinforcement Learning (RL)~\cite{sutton2018reinforcement}, and Natural Language Processing-based (NLP-based)~\cite{manning1999foundations} models.
Supervised learning includes all ML techniques that rely on classification or ranking models of test cases, such as support vector machine, random forests, regression tree, boosting trees, and neural networks.
Unsupervised learning includes all techniques that rely on clustering test cases, such as k-means, expectation-maximization, and agglomerative hierarchical clustering.
Reinforcement Learning (RL) includes techniques that map the TSP problem to an RL problem and use RL algorithms (mainly Q-learning~\cite{sutton2018reinforcement}) to train an RL agent capable of ranking test cases.
NLP-based techniques are used for processing textual data, such as topic modeling, Doc2Vec, and LSTM. NLP-based techniques can also be mixed with other ML or non-ML techniques.

The RL, UL, and NLP-based studies focus on both Test case Prioritization (TP) and Test case Selection (TS), whereas SL studies focus only on TP.
Note that some work (e.g.,~\cite{spieker2017reinforcement}) applies test case selection via executing the top $n\%$ of the prioritized test cases.
We consider a paper to be ML-based TSP if an ML model is used as part of the process for performing TSP, including data preprocessing.
In the following, we discuss the general principles and motivations for each group, along with the similarities and differences across papers within groups.  

\begin{table}[hbt!]
\centering
\caption{Classification of the Selected Papers Based on the Applied ML Techniques}
 \resizebox{0.9\textwidth}{!}{
 \begin{tabular}{llrcc}
 \hline
 ML Technique & References &  (\#, \%)  & TS & TP \\
 \hline
RL         & \cite{spieker2017reinforcement,bertolinolearning,do2020multi,lima2020learning,lima2020multi,rosenbauer2020xcs,shi2020reinforcement} & 7, 24\%  & Yes & Yes\\
UL & \cite{almaghairbe2017separating,carlson2011clustering,chen2011using,kandil2017cluster,khalid2019weight,wang2012using,yoo2009clustering} & 7, 24\% & Yes & Yes\\
SL    & \cite{bertolinolearning,busjaeger2016learning,chen2018optimizing,hasnain2019recurrent,jahan2019version,lachmann2016system,mahdieh2020incorporating,mirarab2008empirical,noor2017studying,palma2018improvement,article,singhmachine,tonella2006using} & 13, 45\% & No  & Yes\\
NLP-based  & \cite{kandil2017cluster,busjaeger2016learning,lachmann2016system,aman2020comparative,medhat2020framework,thomas2014static} & 6, 21\% & Yes  & Yes\\
 \hline
\end{tabular}
}
\label{tab:techniques}
\end{table}

\subsubsection{RL-based Test case Selection and Prioritization}
Recent studies have used RL techniques for TSP, mainly in the context of Continuous Integration (CI). The main motivation of these studies is to benefit from the capacity of RL to seamlessly adapt to the dynamic nature of CI (frequent changes in systems and test suites) and integrate new data into already constructed models without retraining them from scratch. RL studies share similar ideas about the creation of an RL environment by replaying CI logs and training an RL agent via interactions with the environment and receiving rewards. The trained RL agent can take a test case and assigns a score that is used to sort (prioritize) test cases. Despite these commonalities, the studies mainly differ according to the (1) used reward functions and (2) ways the agent learns the optimal policy (policy/value model). Depending on the underlying RL algorithms, RL techniques train an ML model to either estimate the actions' values (value model) and then select the optimal action based on such values, or predict the optimal action directly (policy model). While the use of Deep Neural Network (DNN) as policy/value models is prevalent, it is also possible to use other simpler ML techniques, such as random forest or regression.  

A more detailed investigation of RL studies shows that three of them focus on devising and evaluating different reward functions. The main consideration in the definition of reward functions is the detection of failures, where the highest reward is obtained when failed test cases are prioritized first. A reward is either an instant reward (given to the agent after assigning a priority to each test case) or a delayed reward (given after assigning priority to all test cases). Rewards are calculated in different ways. For example, Spieker et~al.~\cite{spieker2017reinforcement} used a failure count reward, test case failure reward, and time-ranked reward, 
whereas Shi et~al.~\cite{shi2020reinforcement} used a weighted reward function based on the entire execution history.

The other two studies focused on evaluating different ML models as policy models. Bertolino et~al.~\cite{bertolinolearning} not only considered a shallow network for policy model but also a multi-layer perceptron and random forest. Also, Rosenbauer et~al.~\cite{rosenbauer2020xcs} used a XCS classifier system (XCS)~\cite{wilson1995classifier} (a rule-based evolutionary machine learning method) as a policy model. Since the number of test case features is normally reasonable, and potentially important features are known and intuitive, using simpler ML models, other than DNNs, can provide an acceptable accuracy at a lower cost in terms of training data and computation time. 

Three studies used a Multi-Armed Bandit (MAB) approach for TP~\cite{do2020multi,lima2020learning,lima2020multi}. Lima et~al.~\cite{do2020multi} used a MAB-based approach to address test case prioritization in a CI context (TCPCI). They considered three different time budgets and eleven case studies to evaluate their approach. Their evaluation results showed that they outperform RL with an ANN policy model by yielding higher performance. Similarly, Lima et~al.~\cite{lima2020multi} compared the performance of MAB with a Genetic Algorithm (GA). Their results suggested that the MAB approach can fare similarly to the GA algorithm in 90\% of the cases in terms of the percentage of faults detected, Root-Mean-Square-Error (RMSE), and Prioritization Time. In another work, Lima et~al.~\cite{lima2020learning} also considered two strategies for TP using MAB in the CI of Highly Configurable
Systems (HCS): the product Variant Test Set Strategy (VTS), which prioritizes the test set for each product variant in the HCS and the Whole Test Set Strategy (WTS), which prioritizes the test set based on all test cases for all product variants. While these two strategies fared similarly in terms of performance, WTS works well when testing new product variants, since it makes use of the historical information collected from the other ones.

Overall, RL techniques can adapt to changes of the systems under test and test suites, which makes them suitable for continuously changing CI environments. However, to gain full advantage of RL, further research is required as discussed below.

\begin{itemize}
    \vspace{-5pt}
    \item Existing studies are partial as many of the state-of-the-art RL algorithms have not been considered, such as deep Q-learning and deep deterministic policy gradient. Thus, a more extensive investigation in the TSP context is required.
    
    \item As we will discuss in RQ2, all RL studies except one rely solely on execution history to train their agent. Thus, training RL agents using a more extensive feature set should be investigated.
\end{itemize}

\subsubsection{Unsupervised Learning-based Test case Selection and Prioritization}
In the unsupervised learning category, only clustering algorithms have been applied to TSP in the primary studies. Clustering is widely used in general to identify and group similar data points into clusters. A wide range of clustering algorithms are available, which mainly differ according to their required configuration parameters (e.g., number of clusters) and methods to calculate the distance between data instances. The core idea of using clustering in the TSP context is that similar test cases, in terms of coverage and other properties, are expected to have similar fault detection capability. Thus, clustering test cases can help identify groups of similar test cases that can be used as guidance for TSP.
For instance, Carlson et~al.~\cite{carlson2011clustering} devised a clustering algorithm based on coverage information, code complexity, and execution history for prioritizing test cases in each cluster. They first clustered test cases based on coverage information only. Then, they prioritized test cases based on coverage information, code complexity, execution history features, separately. Then, they performed the prioritization using a combination of code complexity and execution history features.
Yoo et~al.~\cite{yoo2009clustering} combined a clustering technique with the Analytic Hierarchy Process (AHP) algorithm. First, they clustered test cases based on their coverage information. Second, they prioritized test cases in each cluster based on the coverage information and expert knowledge and mark the test case with the highest priority in each cluster as the cluster's representative. Third, they performed inter-cluster prioritization based on their representative test cases. Finally, they performed the final prioritization of all test cases from clusters in a circular order in which the clusters are placed based on their assigned priorities from the third step.

Four of the UL papers~\cite{chen2011using,kandil2017cluster,khalid2019weight,wang2012using} used the K-means algorithm, which requires the number of clusters ($k$) to be provided upfront.
Three of the UL papers~\cite{almaghairbe2017separating,carlson2011clustering,yoo2009clustering} used a hierarchical clustering algorithm that also requires the number of clusters as input.
The selection of a suitable $k$ in both algorithms requires extensive experiments and tuning.
One paper~\cite{almaghairbe2017separating} considered three different clustering algorithms, which are Expectation-maximization (EM), Density-based spatial clustering of applications with noise (DBSCAN), and Agglomerative hierarchical clustering.

Regarding how the distance between instances is calculated, four studies~\cite{almaghairbe2017separating,carlson2011clustering,chen2011using,wang2012using} used Euclidean Distance and two studies~\cite{kandil2017cluster,yoo2009clustering} relied on the Hamming Distance. The Euclidean Distance is calculated by the sum of squared differences between two vectors. The Hamming Distance is the number of mismatches in the corresponding positions between two vectors of the same length. The Hamming Distance is mainly used for comparing binary data strings. The two papers use coverage information for a test case as binary strings, where each bit indicates whether or not a source code element (e.g., function) is covered by a test case.

Despite the authors' claims about the benefits of clustering in the TSP context, we argue that the practicality of these results is questionable for a number of reasons:

\begin{itemize}
    \item UL techniques, and clustering algorithms in particular, require expensive tuning, regarding distance metrics and the number of clusters, which affects their effectiveness.

    \item The computational complexity of UL techniques is high as, for most of them, finding the optimal solution is an NP-hard problem. This can cause scalability issues for dealing with TSP in a large software system with many test cases. 
\end{itemize}

\subsubsection{Supervised Learning-based Test case Selection and Prioritization}
The majority of the selected papers used supervised learning techniques and addressed TSP as a ranking problem. In particular, these techniques are often guided by three different \textit{ranking models} for information retrieval~\cite{li2011learning}: \textit{pointwise}, \textit{pairwise}, and \textit{listwise} ranking. \textit{Pointwise} ranking takes the features of a single test case and uses a prediction model to provide a relevance score for this test case. The final ranking is achieved by simply sorting the test cases according to these predicted scores. \textit{Pairwise} ranking orders a pair of test cases at a time. Then, it uses all the ordered pairs to determine an optimal order for all test cases. \textit{Listwise} ranking considers a complete list of test cases at once and assigns a rank to each test case relative to other test cases. 

Developing ranking models is a very well-established research field and applying state-of-the-art ranking models is an obvious choice for TSP. 
Our investigation reveals that three papers~\cite{bertolinolearning,lachmann2016system,tonella2006using} used \textit{pairwise} ranking models. More specifically,
Bertolino et~al.~\cite{bertolinolearning} used a state-of-the-art ranking library~\cite{Ranklibs} and evaluated the effectiveness of Random Forest (RF), Multiple Additive Regression Tree (MART), L-MART, RankBoost, RankNet, Coordinate ASCENT (CA) for TP.
Their results show that (MART) is the most accurate model, which is a \textit{pairwise} ranking model.
Lachmann et~al.~\cite{lachmann2016system} used SVM Rank~\cite{SVMRank}, which returns a ranked classification function based on training inputs and is capable to handle input vectors of large size.
Tonella et~al.~\cite{tonella2006using} evaluated Rankboost~\cite{freund2003efficient} for test case prioritization. It requires developers to provide rankings of test case pairs that are used along with statement coverage and cyclomatic complexity for the training of a ranking model.
In addition, \textit{listwise} ranking models were used by two papers~\cite{bertolinolearning,busjaeger2016learning}. In particular,
Busjaeger and Xie~\cite{busjaeger2016learning} used SVM MAP~\cite{yue2007support}, which ranks test cases by training a model based on training data labeled as `relevant' or  `non-relevant'.
Finally, ten papers~\cite{bertolinolearning,chen2018optimizing,hasnain2019recurrent,jahan2019version,mahdieh2020incorporating,mirarab2008empirical,noor2017studying,palma2018improvement,article,singhmachine} used \textit{pointwise} ranking models. Specifically, 
XGBoost~\cite{chen2018optimizing},
RNN~\cite{hasnain2019recurrent},
ANN~\cite{jahan2019version},
NN~\cite{mahdieh2020incorporating},
Bayesian Network~\cite{mirarab2008empirical},
Logistic Regression~\cite{noor2017studying,palma2018improvement},
KNN~\cite{article}, and
SVM~\cite{singhmachine}
were used to provide priority scores or probability of failure for ranking test cases.

Overall, TP based on supervised learning can benefit from state-of-the-art techniques and reach good accuracy. The majority of reported models used ML techniques that are restricted to the classical batch setting and assume the full data set is available before training, do not enable incremental learning (i.e., continuous integration of new data into already constructed models) but, instead, regularly reconstruct new models from scratch. This is not only very time-consuming but also leads to potentially outdated models, which can be problematic, especially in CI environments. For example, the MART ranking model is an ensemble model of boosted regression trees. Boosting algorithms, as a class of ensemble learning methods, are designed for static training, based on a fixed training set. Thus, they cannot be directly and easily applied to online learning and incremental learning~\cite{zhang2019incremental}. Supporting incremental learning in boosting algorithms is an active research area for which no solution is currently available in existing libraries~\cite{Ranklibs}. This causes practical issues, since modeling performance gradually decays after some cycles, and a new model needs to be trained based on the most recent data.

\subsubsection{NLP-based Test case Selection and Prioritization}
The use of NLP is limited to six papers. 
The core motivation to apply NLP techniques is to exploit information in either textual software development artifacts (e.g., defect description) or source code that is treated as textual data.   
Thomas et~al.~\cite{thomas2014static} relied on an NLP topic modeling technique to transform test cases into vectors, then calculated the distance between pairs of test cases. They then prioritized test cases by maximizing their distances to already prioritized test cases using a greedy algorithm. 

Aman et~al.~\cite{aman2020comparative} used three different kinds of NLP techniques, including topic modeling, Doc2Vec (PV-DM), and Doc2Vec (PV-DBoW) to vectorize test cases, and then calculated the distance between pairs of test cases using three different distance metrics (Manhattan distance, Euclidean distance, and angular distance). The highest priority was then assigned to the test case which was farthest to others. Then, they prioritized the remaining test cases based on their distance from the set of already prioritized test cases.

Medhat et~al.~\cite{medhat2020framework} used NLP to preprocess the specifications that describe the components and technologies of the system under test, then ran the LSTM~\cite{LSTM} algorithm to classify these specifications into four standard components (user device, protocols, gateways, sensors, actuators, and data processing). Based on this classification, test cases that belonged to these standard components were selected. Then, they used search-based approaches (genetic algorithms and simulated annealing) to prioritize the selected test cases.

Lachmann et~al.~\cite{lachmann2016system} created a dictionary of commonly occurring words in the description of all test cases. Then, they used NLP techniques, i.e., tokenization, filtration, and stemming, to preprocess test case descriptions before they transformed them into vectors based on word occurrences. They prioritized test cases by employing a ranked Support Vector Machine (SVM) using the transformed textual data in addition to coverage information and execution history.

Overall, despite its potential, the current use of NLP in a TSP context is very limited. Further research in this context is essential, two examples of which are discussed in the following.

\begin{itemize}
    \item \textit{Coverage analysis.} Static and dynamic coverage analysis techniques are core elements of non-ML TSP techniques. However, both techniques are computationally expensive. As a result, recently, researchers have tried to use information retrieval techniques to compute coverage based on the similarity of either test cases with each other or test cases with source code elements (i.e., file, class, or method). The former helps find similar test cases in terms of coverage that can then be used as guidance for TSP. The latter assumes similarity to be a coverage indicator and, therefore, selects test cases that are similar to changed source elements to achieve coverage. In future research, NLP techniques should be further investigated to devise more effective similarity-based TSP techniques. 
    
    \item \textit{Feature extraction.} Many textual software development artifacts are a rich source of information to devise useful features for training more effective ML models for TSP. Also, recent NLP-based techniques for vectorizing code (e.g.,~\cite{alon2019code2vec}) could help extract new potentially important features from source code. In both cases, the application of NLP techniques appears to be a promising and interesting research direction.
\end{itemize}

\mybox{\textbf{\textit{RQ1 Conclusion:}} ML-based TSP mainly employs RL, UL, SL, and NLP-based models. Unlike RL-based TSP techniques, UL-based and SL-based TSP techniques are unsuitable in CI contexts, since they require the reconstruction of models from scratch on a regular basis. Future research should further investigate the use of NLP techniques to devise more effective similarity-based TSP techniques.}

\subsection{\textbf{RQ2.} What are the features used by ML-based TSP techniques?}
Figure~\ref{fig:features} shows a high-level hierarchy of features that are used across selected papers. Also, Table~\ref{tab:featuresets} summarizes what papers rely on what features. At the highest level, we classify the features into five groups, as discussed in the following. 

\begin{itemize}
    \item \textbf{Code complexity.} These features refer to metrics that are used to indicate the complexity of source code entities (file, method, class, code snippet). A comprehensive list of complexity metrics is presented in~\cite{nunez2017source}. Overall, ten of the selected papers used code complexity metrics. While a wide range of complexity metrics are available, only five~\cite{bertolinolearning,mahdieh2020incorporating,mirarab2008empirical,article,singhmachine} out of the ten papers used a comprehensive list of metrics, and the remaining papers mainly employed Lines of Code (LoC), since it can be easily computed. Indeed, the collection of most of the complexity metrics requires static analysis techniques that render their application challenging in many CI contexts. 
    Although Jahan et al.~\cite{jahan2019version} aimed at accounting for the number of requirements associated with each test case, they assumed that the number of conditional statements covered in test cases captures indirectly requirement coverage. Hence, we classify this feature as code complexity.
    
    \begin{figure}
        \centering
        \includegraphics[width=15cm]{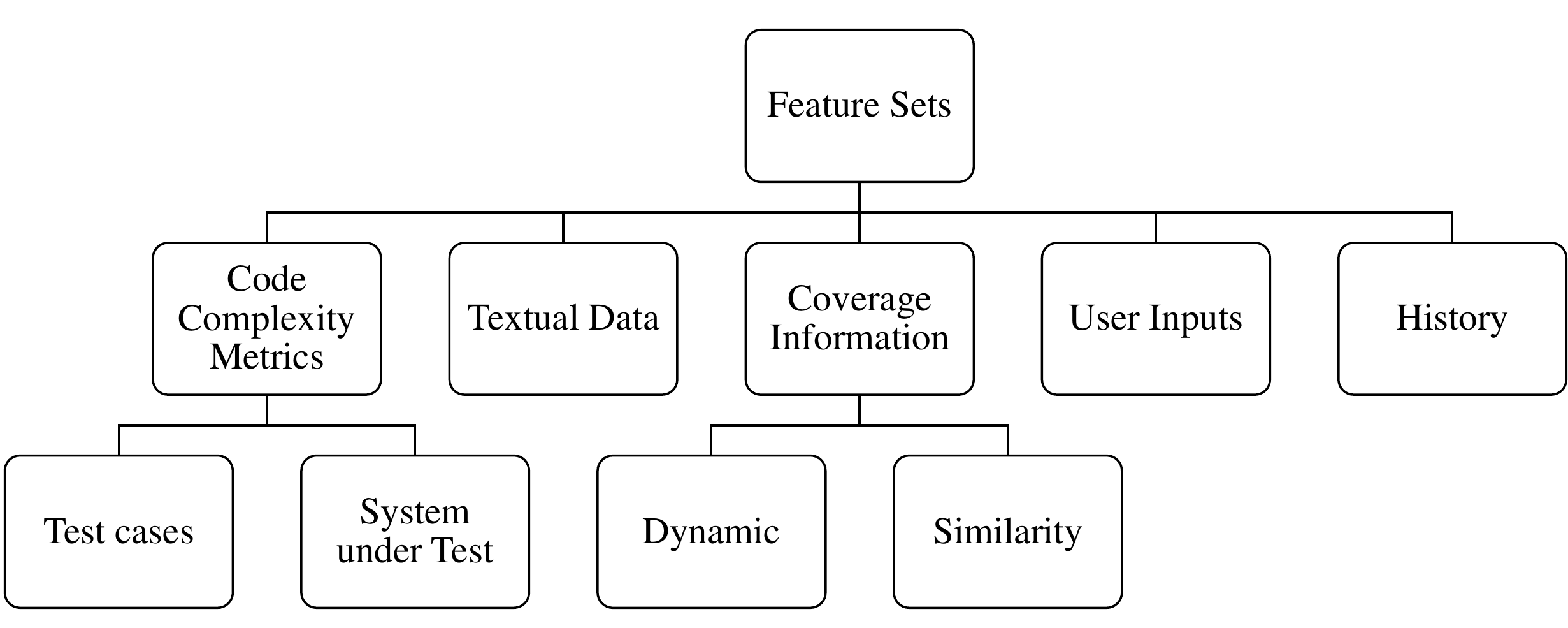}
        \caption{High-level Hierarchy of Features Used by the Selected Papers }
        \label{fig:features}
    \end{figure}

\begin{table}
\centering
\caption{Feature Sets for ML-based TSP Techniques}
 \resizebox{1\textwidth}{!}{
 \begin{tabular}{lp{9cm}p{2.3cm}}
 \hline
    Feature Set & Description & References \\\hline
    
    Code complexity  & A subset of complexity metrics~\cite{nunez2017source}  & \cite{bertolinolearning,carlson2011clustering,jahan2019version,mahdieh2020incorporating,mirarab2008empirical,noor2017studying,palma2018improvement,article,singhmachine,tonella2006using} \\ \hline

    Textual data & Identifiers names, comments and string literals that are extracted from source code & \cite{almaghairbe2017separating,busjaeger2016learning,lachmann2016system,aman2020comparative,medhat2020framework,thomas2014static,kandil2017cluster} \\ \hline

    Coverage information & number of methods exercised by a test case; Production code covered by a test case (e.g., in terms of LOC); Number of system requirements covered by a test case  & \cite{carlson2011clustering,chen2011using,kandil2017cluster,wang2012using,yoo2009clustering,busjaeger2016learning,chen2018optimizing,jahan2019version,lachmann2016system,mahdieh2020incorporating,mirarab2008empirical,noor2017studying,palma2018improvement,tonella2006using,medhat2020framework} \\ \hline

    Execution history & Last execution time (duration) of test cases; Number of failed tests in the current commit; Number of failed tests per test class $n$ commits before the current one; Total execution time of all the tests of the test class; Last time the test class was run; Average execution time; Test cases' verdict; Last $n$ test verdicts which refer to the history length; Number of faults detected by each test case; Method execution of test case; Test age; Failure age; Failure priority  & \cite{spieker2017reinforcement,bertolinolearning,do2020multi,lima2020learning,lima2020multi,rosenbauer2020xcs,shi2020reinforcement,carlson2011clustering,busjaeger2016learning,chen2018optimizing,hasnain2019recurrent,lachmann2016system,noor2017studying,palma2018improvement,medhat2020framework} \\\hline
    
    User inputs  & Requirement priority assigned by the test manager; Priority of test case pairs assigned by users &\cite{khalid2019weight,yoo2009clustering,tonella2006using} \\ \hline
 \end{tabular}
} 
\label{tab:featuresets}
\end{table}

    While it is possible to compute such metrics based on the test code or the source code of the system under test, many papers exclusively focused on the former. The main motivation for using code complexity metrics is based on the hypothesis that more complex code has a higher probability of failure and tends to have longer execution times. Yet, relying only on complexity metrics of the test code is not enough, since the main goal of TSP is to detect faults early in the production code.

     \item \textbf{Textual data.} As discussed earlier, many textual software development artifacts are a rich source of information to devise useful features for training more effective ML models for TSP. Seven papers~\cite{almaghairbe2017separating,busjaeger2016learning,lachmann2016system,aman2020comparative,medhat2020framework,thomas2014static,kandil2017cluster} relied on textual data. Such data appears to be a promising and interesting research direction to devise more accurate TSP techniques, especially when access and analysis of the source code is not an option. 
    
    \item \textbf{Coverage information.} 
    These features capture the coverage of test cases either as covered source code elements (i.e., file, class, and method) or requirements.
    Source code coverage can be collected using three methods: dynamic analysis, static analysis, and code similarity analysis. Dynamic analysis is the most precise method but requires executing tests on an instrumented system version (i.e., binary or source code instrumentation). Dynamic analysis is platform-dependent and time-consuming for large codebases, and may not be applicable to systems with real-time constraints, since code instrumentation may cause timeouts or interrupt normal test execution. Static analysis makes it easier to collect coverage information and conduct impact analysis and have therefore drawn more attention. However, it tends to overestimate coverage and change impact, affecting the accuracy of TSP methods. Finally, code similarity analysis relies on NLP and information retrieval techniques to analyze coverage based on source code similarity between the test cases and system elements. Coverage features are used by fifteen papers, eight of which~\cite{carlson2011clustering,chen2011using,wang2012using,yoo2009clustering,mahdieh2020incorporating,mirarab2008empirical,palma2018improvement,medhat2020framework} using dynamic coverage analysis techniques to extract the number of covered methods or lines of code, four of which~\cite{kandil2017cluster,jahan2019version,lachmann2016system,tonella2006using} using static coverage analysis, three of which~\cite{busjaeger2016learning,chen2018optimizing,noor2017studying} using both dynamic and static coverage analyses.
    
    Also, Busjaeger and Xie~\cite{busjaeger2016learning} calculated the coverage score for a given test case based on the similarity between its code and changed files as well as their file paths. More specifically, for a given changed file and test case, they first extracted words from the file-system path and the textual content of the changed file (only changed parts of the file) and the test case. They then calculated the similarity score for the test case with respect to the changed file as the cosine similarity between words in the changed file and test case, respectively.
    
    Requirements specifications describe of the functional and non-functional requirements of a software system to be developed. There are four papers~\cite{jahan2019version,kandil2017cluster,lachmann2016system,medhat2020framework} using coverage based on requirements. Jahan et~al.~\cite{jahan2019version} relied on the number of changed requirements represented by the number of conditional statements changed in test cases. Kandil et~al.~\cite{kandil2017cluster} used the number of modules covered by user stories. Lachmann et~al.~\cite{lachmann2016system} and Medhat et~al.~\cite{medhat2020framework} focused on the number of requirements covered by test cases.

    \item \textbf{Execution history.} These features capture information about previous test case executions, including test execution time, test results or verdict (i.e., failed/passed), age of test cases, and aggregate features such as average execution time or failure rate. In general, thanks to test automation tools, specifically in a CI context, historical features are easy to collect. Thus, most of the papers (15 papers) used such features.
    Due to the frequent execution of regression test suites, the volume of execution history data is continuously and quickly growing. Thus, papers often limit the use of such data to the last $n$ test executions, which is referred to as the \textit{history length}. 
    
    \item \textbf{User input.} These features, such as test case priority, are specified by users. The data collection for these features requires manual effort and that makes them less practical, especially in a CI context and for large software projects. Only three of the papers relied on user inputs~\cite{khalid2019weight,yoo2009clustering,tonella2006using}. Khalid et~al.~\cite{khalid2019weight} assumed that the main goal of test case prioritization is to effectively use time and budget to execute the highest priority test cases first based on customer preferences. Thus, they proposed a technique in which they use a customer-assigned priority to each test case extracted from business requirements. Also, Tonella et~al.~\cite{tonella2006using} assumed that test engineers usually know the relative priority of test cases. Thus, they proposed a test case prioritization technique that takes advantage of such user knowledge through a machine learning algorithm called case-based ranking (CBR), in which test engineers specify relevance priorities of test cases in the form of \textit{pairwise} comparisons. 
\end{itemize}

In addition, some features are specific to certain application contexts. For example, Sharma et~al.~\cite{article} used page name, number of buttons, number of toggles, number of hyperlinks, and the number of images as features in the context of web application testing.
Hasnain et~al.~\cite{hasnain2019recurrent} used response time and throughput metrics by calculating the amount of data transferred in kilobytes as well as web service latency as features in the context of Internet of Things (IoT) systems.

We further analyzed the papers with respect to data collection processes and tools. All papers, except the ones that used open-source datasets
briefly addressed the data collection process. However, none of them except one~\cite{mahdieh2020incorporating} reports practical challenges in the calculation of coverage data, the time complexity of coverage-based techniques, and the computation time of the coverage techniques based on different case studies.   

Overall, existing ML-based TSP techniques did not take full advantage of potentially available and useful data and, therefore, potentially relevant features remain unused. The features used by most of the papers were only limited to the features that are computed based on easily accessible data. For instance, (1) four of the RL papers used three available data sets containing only execution history features, (2) most of the papers used LoC as the only code complexity feature, (3) coverage information was mainly used by papers having Java and C subjects due to available tooling support for source code and byte-code instrumentation (e.g., JaCoCo~\cite{JaCoCo}, Clover~\cite{Clover}, and GCov~\cite{GCOV}). Given the above limitations, interesting research directions include:

\begin{itemize}
    \item  Investigating the use of a more comprehensive set of potentially relevant features for ML training and perform a trade-off analysis on the relative benefits and costs of features.
    
    \item  Performing a thorough and systematic evaluation of existing tools and solutions to address TSP data collection needs. Requirements for data collection tools specifically addressing TSP should be identified and addressed with new tools when needed. For example, there is lack of tools for automatically analyzing the build logs of popular CI tools (e.g., Travis~CI\footnote{\url{https://www.travis-ci.com}}) and extracts the test case execution details (execution time and test case verdicts) required for TSP. Existing tools, such as the build log analyzer used for TravisTorrent~\cite{msr17challenge}, only support the analysis of the logs at the build level, i.e., the execution time, coverage, and results of individual test cases are not captured.
\end{itemize}

\mybox{\textbf{\textit{RQ2 Conclusion:}} ML-based TSP techniques tend to rely on features that are easy to collect or compute. Besides those features, future research should look into a more comprehensive set of potentially relevant features (e.g., CI-related features, such as build configuration complexity, runtime environments, timeouts, etc.) for ML training and investigate their benefits, costs, and the tools to collect them.}

\subsection{\textbf{RQ3.} How are ML-based TSP techniques evaluated?}
\label{RQ3}

In this section, we discuss the evaluation metrics as used and reported in the selected papers, their calculation methods, and their strengths and drawbacks. Then, we describe the subjects based on which the studies are evaluated and assess how adequate they are to draw conclusions on the practical effectiveness of TSP techniques.

\subsubsection{Evaluation Metrics}
Eight metrics have been used for evaluating ML-based TSP techniques, which can be categorized into two groups: specific and general. The former includes Average Percentage of Fault Detection (APFD)~\cite{elbaum2002test} and its extensions, such as Normalized APFD (NAPFD)~\cite{qu2007combinatorial} and cost-cognizant APFD (APFDc)~\cite{elbaum2001incorporating}. These metrics are specifically defined to measure the performance of TSP techniques based on how early they can detect faults. The latter includes general classification performance metrics (i.e., accuracy, precision, recall, and $\text{F1}$-score) and Rank Percentile Average (RPA) that measures how close is a predicted ranking to an actual ranking, regardless of the application context.

\paragraph{Specific Evaluation Metrics for TSP\\}~

\noindent APFD is the most widely used evaluation metric for TP. It evaluates the effectiveness of a TP technique via measuring the area under the curve when plotting the proportion of executed test cases on the x-axis and the proportion of detected faults on the y-axis~\cite{elbaum2002test}. It ranges from $0$ to $1$, with a higher number implying a faster fault detection. APFD can also be calculated based on the number of test case failures when the number of faults detected in a test suite is not available.

Let $T$ be a test suite of $n$ test cases, the \textit{APFD} of a prioritized test suite $T'$ is calculated using the following formula:
$$
APFD= 1 - \frac {\sum_{i=1}^{m}TF_i} {nm} + \frac{1}{2n}
$$ 
where $m$ refers to the total number of faults, $n$ refers to the total number of test cases, $TF_i$ returns the rank of the first test case that reveals the $i$th fault in test suite $T'$. 

APFD treats all test cases equally in terms of cost and fault severity, which is not always realistic. To address this issue, APFDc was introduced~\cite{elbaum2001incorporating} to take test execution cost and fault severity into account~\cite{elbaum2001incorporating}. 
The formula of APFDc is defined as follows:
$$
APFDc= \frac{\sum_{i=1}^{m}(f_i * (\sum_{j=TF_i}^{n}t_j-0.5*t_{TF_i}))}{\sum_{i=1}^{n}t_i*\sum_{i=1}^{m}f_i}
$$

where $t_i$ refers to the execution cost of the $i$th test case, and $f_i$ denotes the severity of the $i$th fault.
Note that two of the studies~\cite{chen2018optimizing,do2020multi} used APFDc as an evaluation metric for TSP. They both assumed that all faults had the same severity and measured costs based on execution times of test cases. They calculated APFDc as follows:

$$
APFDc= \frac{\sum_{i=1}^{m}(\sum_{j={TF_i}}^{n}t_j-0.5*t_{TF_i})}{\sum_{j=1}^{n}t_j*m}
$$

APFD assumes that the ranked test cases can detect all detectable faults, 
i.e., all ranked test cases are to be executed. However, in TSP, the test execution budget might be limited to only the top $n\%$ of test cases (i.e., not the entire test suite is executed). In such scenarios, variations in the test suite are likely to occur in each run, which might not detect all faults, making APFD unsuitable in practice.
Thus, an extension of APFD was introduced called Normalized APFD (NAPFD)~\cite{qu2007combinatorial} to address this issue by using both the ratio between detected and total faults within the prioritized test suite $(p)$ as follows:

$$
NAPFD= p - \frac {\sum_{i=1}^{m}TF_i} {nm}  + \frac{p}{2n}
$$

Note that when all test cases are executed and all faults are detected (i.e.,  $p = 1$), as expected, NAPFD is equal to APFD.

Further, all APFD metrics assume that faults occur sufficiently frequently to provide an accurate assessment. However, with the introduction of CI, where software developers may build the system several times a day, the system quality is being continuously improved and there are many builds without or with only a few faults. When there are only a few faults, a small difference in fault detection across the ranking (e.g., not detecting one fault) can have a large impact on the APFD value. Also, the calculation of APFDs is obviously impossible when there is no fault. In this case, studies often assume APFD values of $1$ or $0$~\cite{spieker2017reinforcement,elbaum2002test}, both of which introducing bias in the evaluation results.

\paragraph{General Evaluation Metrics for TSP\\}~ 

\noindent\textbf{Accuracy, precision, recall, and $\text{F1}$-score.}
Assuming a classifier that predicts whether an item (e.g., test case) belongs to a certain class or not (returns true for positive prediction and false otherwise), we refer to a classification of an item as: 

\begin{itemize}
    \item \textit{True Positive} (TPo) when the item actually belongs to the class and prediction is \textit{true},
    \item \textit{False Positive} (FPo) when the item actually does not belong to the class and prediction is \textit{true},
    \item \textit{True Negative} (TNe) when  the item actually does not belong to the class and prediction is \textit{false}, or
    \item \textit{False Negative} (FNe) when the item actually belongs to the class and prediction is \textit{false}
\end{itemize}

TPo and TNe refer to correct classification cases (ground truth and classifications are identical), whereas FPo and FNe refer to incorrect classifications. According to the above definitions, Table~\ref{tab:genmetrics} shows how the accuracy, precision, recall, and $\text{F1}$-score metrics are defined.

\begin{table}[hbt!]
    \centering
    \caption{Description of Accuracy, Precision, Recall, and $\text{F1}$-score Metrics}
    \begin{tabular}{ p{1.5cm}p{5.5cm}p{5.5cm}}
     \hline
     Metric &  Description& Formula\\
     \hline
     Accuracy &   The ratio of correctly predicted samples to total observations & 
     \[ 
     \text{accuracy} = \frac{\text{TPo + TNe}}{\text{TPo + TNe + FPo + FNe}}  
    \]\\  \hline
    Precision & The ratio of TPo to the total predicted positive &
    \[
        \text{precision} = \frac{\text{TPo}}{\text{TPo + FPo}} 
    \] \\  \hline
    Recall & The ratio of True Positive to the total actual positive &
    \[
        \text{recall} = \frac{\text{TPo}}{\text{TPo + FNe}}.
    \]\\  \hline
    $\text{F1}$-score & The ratio between precision and recall &
    \[ \text{F1}=2 \times \frac{\text{precision} \times \text{recall}}{\text{precision} + \text{recall}}   
    \]\\
     \hline
    \end{tabular}
    \label{tab:genmetrics}
\end{table}

\noindent\textbf{Rank Percentile Average (RPA).} The RPA metric was proposed by Bertolino et~al.~\cite{bertolinolearning} to compute how close a predicted ranking was to the optimal ranking. It assumes the priority scores of the test cases to be an increasing integer from 1 to $k$, where $k$ is the number of test cases to prioritize. The higher the score, the higher the priority. Given that $r_i = i$ is the actual ranking score of test case $i$, RPA is calculated as follows:

 \begin{equation}
 \text{RPA} = \frac{1}{k}\sum_{m=1}^{k} \frac{1}{r}  \sum_{i=k-m+1}^{k} r_i = \frac{\sum_{m=1}^{k} \sum_{i=k-m+1}^{k} r_i}{k^2(k+1)/2}.
 \end{equation}
 
 The maximum value of RPA is reached when the predicted ranking is equal to the optimal ranking, which can be defined as $RPA_M$:
 
    $$
    RPA_M= 1 - \frac{\sum_{i=1}^{k-1}(k-i)(k-i+1)}{k^2(k+1)}
    $$

Then the Normalized-Rank-Percentile-Average (NRPA), which ranges from 0 to 1, is defined as follows:

    $$
    NRPA= \frac{RPA}{RPA_M}
    $$

Discussing the general (dis)advantages of evaluation metrics is not the focus of our study. However, in a TSP context, these evaluation metrics in their default form can be misleading in some cases, especially when attempting to account for the fault detection capability of TSP techniques. More specifically, the quick detection of faults is considered important in most software systems but is ignored by general evaluation metrics in their default form, as they treat all test cases equally, disregarding their verdicts. This can lead to situations in where (a) a TS technique (TS$_1$) has higher general metric scores compared to another TS technique (TS$_2$), but may actually detect fewer faults than TS$_2$ or (b) a TP technique (TP$_1$) with higher general metric scores compared to a TP technique (TP$_2$), may detect faults later than TP$_2$.

To make the above argument clearer, assume that we have eight test cases $T_1 \dots T_8$ with an identical execution time and include three failed test cases and five passed test cases. Table~\ref{tab:tsemetrics} shows an example of two TS techniques where higher accuracy, precision, and F1-score values are given to techniques that detect fewer faults. In these cases, one should pay more attention to the recall metric, which is defined as the proportion of selected failed test cases among all failed test cases, since the goal is to detect as many faults as possible~\cite{busjaeger2016learning}.

\begin{table}[ht!]
\centering
\caption{The Output of Two Different TS When Test Cases are $\langle$0 0 0 1 1 0 0 1$\rangle$, Indicating T$_4$, T$_5$, and T$_8$ are Failed Test Cases. Selecting a Failed Test Case is Considered a True Positive Case.}
  \begin{tabular}{ p{4.5cm}p{1.5cm}p{1.5cm}p{1.5cm}p{1.5cm}  }\hline
         Test Case  & Accuracy  &  Precision  &  Recall &  $\text{F1}$-score \\  
 \hline
   TS$_1$ = $\langle$0 0 0 1 0 0 0 0$\rangle$ & 0.75  & 1.00  & 0.33  & 0.50\\
   TS$_2$ = $\langle$1 1 0 1 1 1 1 0$\rangle$ & 0.38   & 0.33 & 0.67  & 0.44 \\
 \hline
 \end{tabular}
 \label{tab:tsemetrics}
\end{table}

Similarly, in a TP context, it is possible for a TP technique that assigns lower ranks to failed test cases (the lower the rank, the lower the priority of execution) to be given a higher NRPA compared with a TP technique that assigns higher ranks to failed test cases. Table~\ref{tab:tp} shows an example of two TPs (TP$_1$ and TP$_2$) for ten test cases with the same execution time. As shown, a higher NRPA is given to TP$_1$ even though it does not rank failed test cases higher compared to TP$_1$. Hence, the NRPA metric is not suitable in these cases.

\begin{table}[ht!]
\centering
\caption{The Output of Two Different TP Techniques When Test Cases are $\langle$0 0 0 1 1 0 1 1$\rangle$, Indicating T$_4$, T$_5$, T$_7$, and T$_8$ are Failed Test Cases.}
  \begin{tabular}{ p{8.5cm}p{1.5cm}}
  \hline
TP  & NRPA \\  
 \hline
TP$_1$ = $\langle$8 7 6 3 2 5 4 1$\rangle$ & 0.96\\
TP$_2$ = $\langle$4 5 7 1 8 2 3 6$\rangle$ & 0.80 \\
\hline
 \end{tabular}
\label{tab:tp}
\end{table}

Moreover, we found that some evaluation metrics are calculated differently in TSP, even when modeling was performed using the same SL or UL technique. 
For example, in a TS context, True Positive (TPo) is defined as the number of selected failed test cases~\cite{almaghairbe2017separating,chen2011using,kandil2017cluster}, whereas in a TP context, TPo is defined as the number of test cases being correctly ordered~\cite{jahan2019version,medhat2020framework} or the number of test cases with the same priority as the actual priority of test cases~\cite{khalid2019weight}. Note that the calculation of TPo was based on test case clustering in all those studies.
Hence, comparing TSP results using these general metrics would be meaningless due to the above differences in definitions.
APFDs, however, were consistently calculated in all studies that used them as an evaluation metric of ML-based TSP techniques, thus making the comparison among studies using APFDs meaningful.

Table~\ref{tab:metrics} summarizes the used ML-based TSP evaluation metrics, along with their context of application (TS, TP, or TSP) and the papers that used them. We observe that the majority of the papers evaluated TP based on APFDs, except for Bertolino et~al.~\cite{bertolinolearning} (NRPA), Medhat et~al.~\cite{medhat2020framework} (precision), Khalid et~al.~\cite{khalid2019weight} (accuracy) and Sharma et~al.~\cite{article} (precision, recall, and $\text{F1}$-score).
Also, three of the papers~\cite{almaghairbe2017separating,chen2011using,kandil2017cluster} used $\text{F1}$-score to evaluate the effectiveness of TS techniques. Busjaeger and Xie~\cite{busjaeger2016learning} used recall for the same purpose.

\begin{table}[ht!]
\centering
\caption{Metrics Used to Evaluate ML-based TSP techniques}
 \resizebox{1\textwidth}{!}{
 \begin{tabular}{p{6cm} p{3.1cm}p{5.2cm}}
 \hline
 Evaluation Metrics & Context & References\\
 \hline
 APFD & TP & \cite{carlson2011clustering,yoo2009clustering,busjaeger2016learning,jahan2019version,lachmann2016system,mahdieh2020incorporating,mirarab2008empirical,tonella2006using,aman2020comparative,thomas2014static}\\
 NAPFD                                     & TP  & \cite{do2020multi,lima2020learning,lima2020multi,rosenbauer2020xcs,shi2020reinforcement}\\
 NAPFD                                     & TSP & \cite{spieker2017reinforcement} \\
 APFDc                                     & TP  & \cite{do2020multi,chen2018optimizing}\\
 Accuracy                                  & TP  & \cite{khalid2019weight,jahan2019version}\\
 Precision                                 & TSP & \cite{jahan2019version,medhat2020framework}\\
 Recall                                    & TSP  & \cite{busjaeger2016learning,jahan2019version}\\
 $\text{F1}$-score                         & TS  & \cite{almaghairbe2017separating,chen2011using,kandil2017cluster}\\
 Precision, Recall, and $\text{F1}$-score  & TP  & \cite{article} \\
 NRPA                                       & TP  & \cite{bertolinolearning}\\
 \hline
\end{tabular}
}
\label{tab:metrics}
\end{table}

Overall, as we discussed, both general and specific metrics have issues and, more importantly, general metrics treat all test cases equally disregarding their verdicts and APFDs can not be calculated when there is no or a few faults. In this context, using weighted versions of general metrics, such as weighted recall, allowing us to give a higher significance to the classification or ranking of failed test cases, can be investigated. 

\subsubsection{Used Subjects}
Analyzing the subjects that are used for evaluation in the primary studies can help us assess the external validity of reported results. Indeed, we can appraise how representative and realistic these subjects are in the TSP context. To address this question, we analyzed the subjects that are made publicly available (126 subjects) based on the following criteria.

\begin{itemize}
    \item \textit{Number of test cases.} The number of test cases can be an indicator of the complexity of the regression testing warranted by systems. In practice, TSP is typically applied to large software systems that tend to have many regression test cases. Hence, evaluating a TSP technique based on a subject with a small number of test cases is often insufficient to demonstrate the practical benefits of the techniques. 
    However, the number of test cases is not the only factor to account for when evaluating a TSP technique. Test execution times are also relevant since they can vary a great deal across test cases.
   
     \item \textit{Average execution time of test cases per build.} One of the main motivations of the TSP techniques is decreasing regression testing time. Also, the computation time of TSP techniques should be significantly less than regression testing time for their application to make sense. Thus, applying TSP techniques to systems whose regression testing takes negligible time may not be practically viable. This criterion concerns the average execution time of regression testing to assess whether the application of TSP on subjects can lead to practical benefits. 

    \item \textit{Number of builds and their failure rate.} To show the effectiveness of TSP techniques and deal with their inherent degree of randomness, they need to be evaluated based on subjects with a sufficient number of builds. Also, since the main goal of TSP is to reveal regression faults, subjects should include a sufficient number of failed builds, i.e., enough builds that reveal at least one regression fault. Using subjects that do not meet these conditions is a threat to the validity of the evaluation results. This is especially the case if the evaluation relies on APFD or defines TPo based on selected failed test cases. Also, having a sufficient number of failed builds is vital for the creation of balanced datasets that are required by many of the ML techniques.
\end{itemize}

Our investigation shows that papers use 126 open source subjects. Table~\ref{tab:datasources} shows the data sources of the subjects used by the studies. We observe that about half of the studies use subjects relying on publicly available datasets, such as Software-artifact Infrastructure Repository\footnote{\url{https://sir.csc.ncsu.edu}} (9 studies), ABB Robotics \& Google Shared Dataset of Test Suite Results (GSDTSR)\footnote{\url{https://bitbucket.org/HelgeS/atcs-data}}
(5 studies), and Defects4J~\cite{just2014defects4j} (3 studies). Ten of the studies extend the available datasets with additional features. However, those datasets are quite old and may not capture the current practices of software development, testing, and CI settings. For example, the most recent update of the SIR dataset was before 2016. Hence, besides computing additional features, future research should collect data on more recent subjects.

\begin{table}
\centering
\caption{Data Sources of the Subjects Used in the Studies}
\resizebox{.8\linewidth}{!}{
\begin{tabular}{ll}
\hline
Data Source	& References\\
\hline
{Software-artifact Infrastructure Repository}                         & \cite{almaghairbe2017separating,chen2011using,wang2012using,yoo2009clustering,mirarab2008empirical,singhmachine,tonella2006using,aman2020comparative,thomas2014static}\\
ABB Robotics \& GSDTSR                                       & \cite{spieker2017reinforcement,do2020multi,lima2020multi,rosenbauer2020xcs,shi2020reinforcement}\\
Defects4J~\cite{just2014defects4j}                                  & \cite{mahdieh2020incorporating,noor2017studying,palma2018improvement}\\
Open Source Repositories (e.g., GitHub \& GitLab)                   & \cite{bertolinolearning,do2020multi,lima2020learning,lima2020multi,chen2018optimizing,hasnain2019recurrent}\\
Private/Local/Industrial Projects                                   & \cite{carlson2011clustering,kandil2017cluster,khalid2019weight,busjaeger2016learning,jahan2019version,lachmann2016system,article,medhat2020framework}\\
\hline
\end{tabular}
}
\label{tab:datasources}
\end{table}

\begin{figure}
    \centering
    \includegraphics[width=0.7\columnwidth]{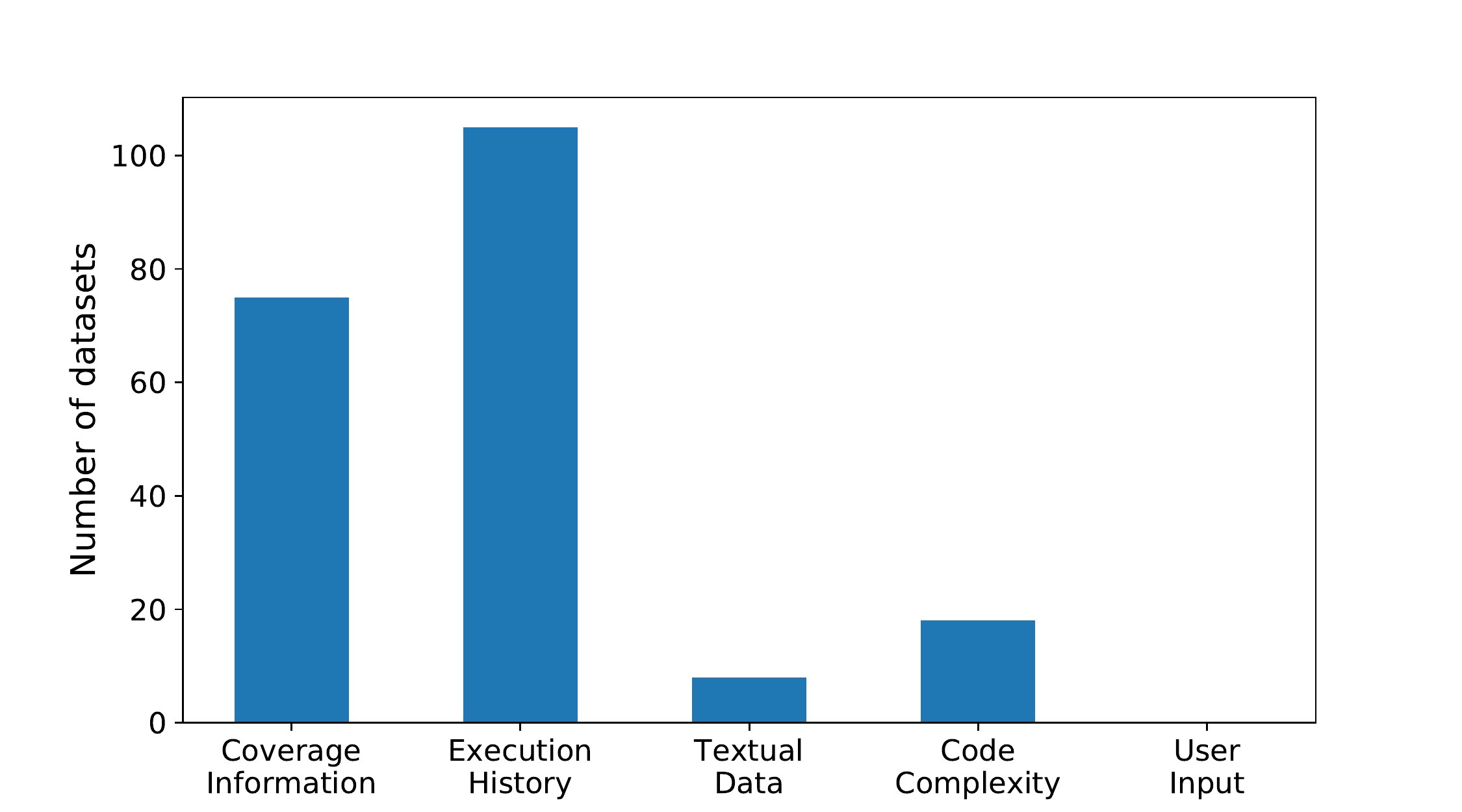}
    \caption{Features Provided by the Used Subjects}
    \label{fig:subjectfeatures}
\end{figure}

\begin{figure}
     \centering
     \begin{subfigure}[b]{0.25\textwidth}
         \centering
         \includegraphics[width=\textwidth]{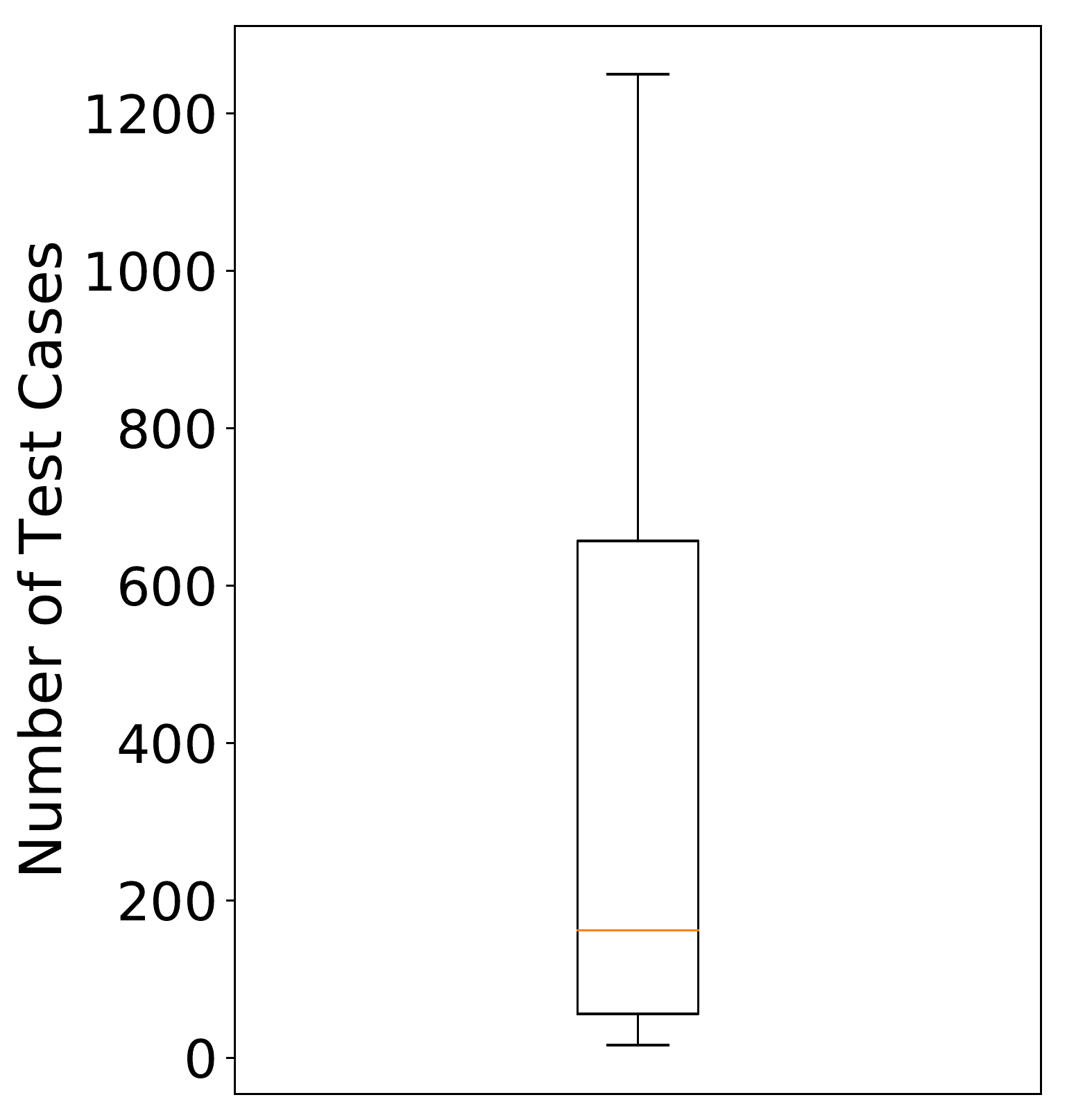}
         \caption{\# of Test Cases}
         \label{fig:test_suite_size}
     \end{subfigure}
     \hfill
    \begin{subfigure}[b]{0.24\textwidth}
         \centering
         \includegraphics[width=\textwidth]{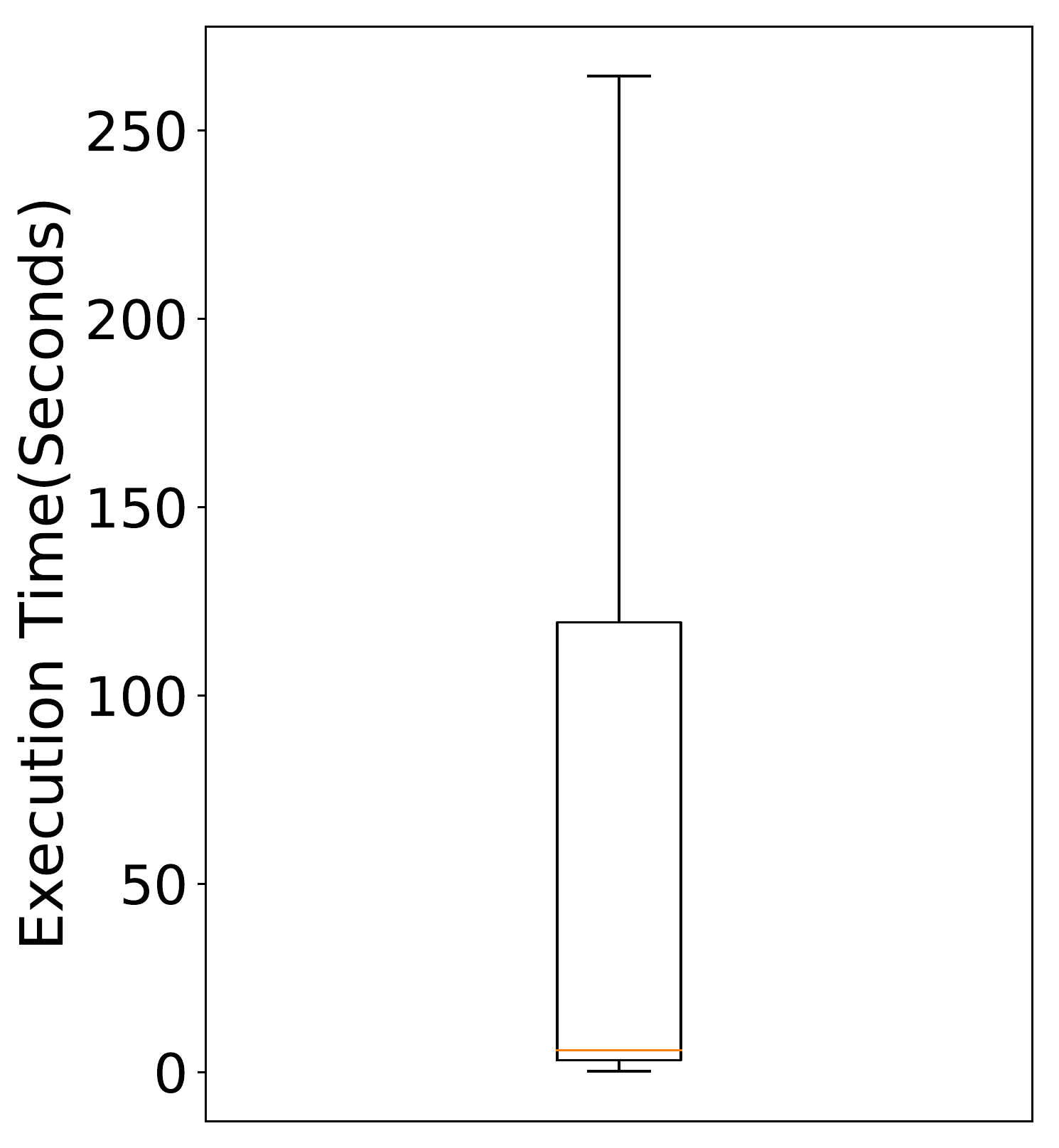} 
         \caption{Testing Time}
         \label{fig:regerssiontime}
     \end{subfigure}
     \hfill
    \begin{subfigure}[b]{0.24\textwidth}
        \centering
        \includegraphics[width=\textwidth]{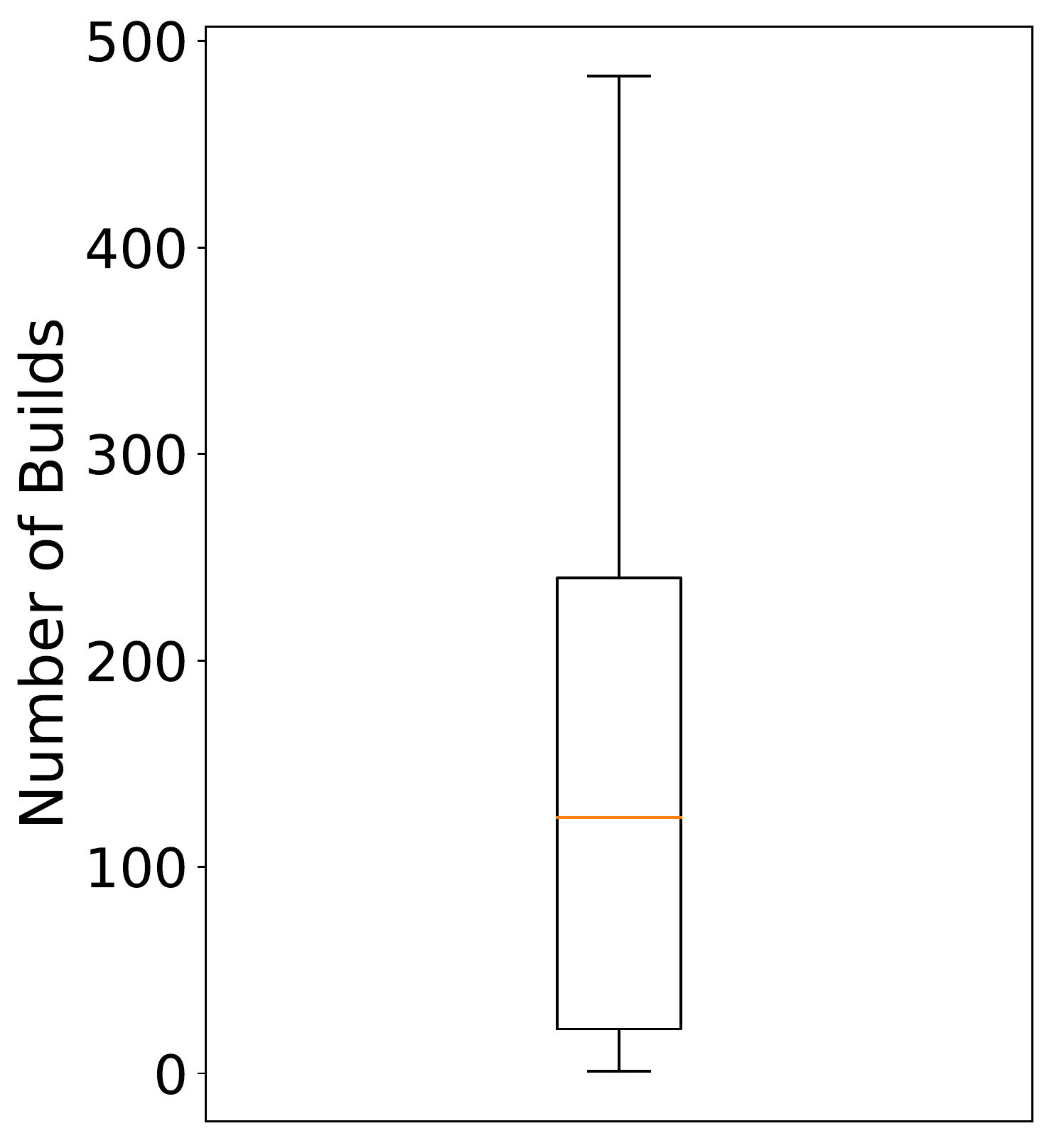}
        \caption{\# of Builds.}
        \label{fig:number_of_builds}
     \end{subfigure}
     \hfill
     \begin{subfigure}[b]{0.229\textwidth}
        \centering
        \includegraphics[width=\textwidth]{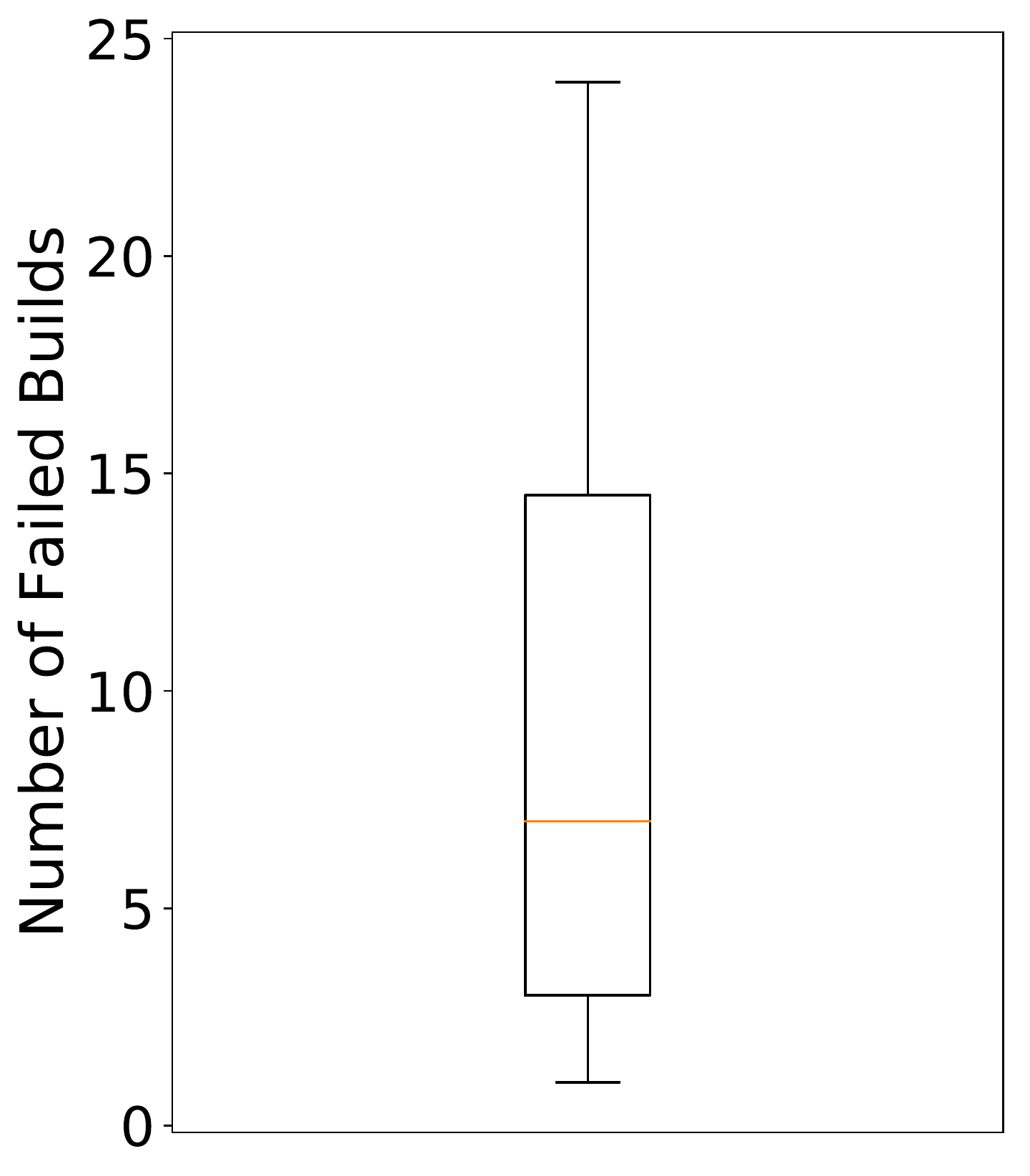}
        \caption{\# of Failed Builds.}
        \label{fig:number_of_failed_builds}
         \end{subfigure}
         \hfill
     \caption{Number of Test Cases, Regressing Testing Time, and Number of (Failed) Builds of Subjects (outliers are removed from all plots)}
    \label{fig:subjects}
\end{figure}

As shown in Figure~\ref{fig:subjectfeatures}, the majority of subjects provide \textit{execution history} (105 subjects) and \textit{coverage information} (75 subjects) features, some of subjects provide \textit{code complexity} (18 subjects), and \textit{textual data} (8 subjects) features.
The \textit{user input} features are not obtained from the codebase or builds of the subjects. Rather, the knowledge about the relative priority of pairs of test cases is obtained from test engineers or subject experts.
In addition, Figure~\ref{fig:test_suite_size} shows the number of test cases associated with subjects. Overall, subjects have a median of 162 test cases, which is large enough to evaluate TSP techniques.  
However, as shown in Figure~\ref{fig:regerssiontime}, which is based on 92 subjects reporting regression testing time, the majority (71\%) of the subjects show a test execution time below 90 seconds, with an overall median of 5.84 seconds.
This result suggests that most of the existing ML-based TSP techniques are evaluated based on subjects for which the application of TSP techniques has no or little practical value. 

Further, Figures~\ref{fig:number_of_builds}~and~\ref{fig:number_of_failed_builds} depict the total and failed numbers of builds based on 40 subjects reporting such data. The number of builds for the majority of subjects (75\%) ranges between 4 and 351 with a median of 150 builds.
Such numbers can be considered sufficient to run extensive experiments and account for randomness in results. However, the number of failed builds, for 50\% of the subjects, is less than 10 with a median of 9.
As discussed above, such a small number of failed builds is a threat to validity, specifically when the evaluation uses APFDc or defines TPo based on failed test cases. This issue was also reported in the context of non-ML-based TSP~\cite{do2006use,luo2018assessing}. To address it, for evaluation purposes, studies focused on non-ML techniques rely on seeded faults, which are typically produced through hand-seeding or mutation fault injection techniques~\cite{do2006use,luo2018assessing}. In the context of ML-based techniques, where the goal is to train an ML model (partly) based on the history of test executions and source code changes, using fault injection techniques is not a valid option since it would introduce random faults into the system that have no relation with its history. 

Overall, based on the above discussions, we argue that most subjects in existing studies cannot be considered adequate for the evaluation of ML-based TSP techniques. This further motivates the need to create data collection tools and an appropriate benchmark for comparing available techniques based on a set of identical and representative subjects. 

\mybox{\textbf{\textit{RQ3 Conclusion:}} Across primary studies, ML-based TSP techniques are evaluated using a variety of metrics. These metrics may not be computed similarly, and treat all test cases equally regardless of their verdicts, making it difficult to compare the results of different techniques. Also, available subjects tend to have extremely low failure rates, making them unsuitable for evaluating ML-based TSP techniques. Future research should also consider more recent subjects with data collected from CI contexts.}

\subsection{\textbf{RQ4.} What is the performance of ML-based TSP techniques?}
\label{RQ4}
The performance results of the ML-based TSP techniques, as reported in the primary studies, including information about their context, feature sets, and evaluation metrics are summarized in Table~\ref{tab:performance_all_papers}.
Table~\ref{tab:baselines} summarizes the comparison baselines used to evaluate the ML-based TSP techniques proposed in the studies. We observe that the majority (nine) of the studies~\cite{spieker2017reinforcement,lima2020learning,busjaeger2016learning,jahan2019version,lachmann2016system,mahdieh2020incorporating,singhmachine,tonella2006using,thomas2014static} compared their proposed techniques with a random approach, in which test cases are ordered randomly, or a cost-only approach, in which tests are ordered based on execution time~\cite{chen2018optimizing}.
In addition, there are three studies~\cite{carlson2011clustering,yoo2009clustering,noor2017studying} in which the evaluation of the proposed techniques was compared to approaches that rank test cases using traditional quality metrics (e.g., fault history or coverage-based) and similarity metrics~\cite{palma2018improvement}.
Other studies~\cite{rosenbauer2020xcs,shi2020reinforcement,chen2011using,hasnain2019recurrent,palma2018improvement} were evaluated in comparison with previous ML-based TSP techniques, such as RETECS~\cite{spieker2017reinforcement}, K-Means~\cite{chen2011using}, RNN~\cite{hasnain2019recurrent}, or regression modeling~\cite{noor2017studying}.
We found no evaluation baselines in eight of the primary studies~\cite{bertolinolearning,almaghairbe2017separating,kandil2017cluster,khalid2019weight,mirarab2008empirical,article,aman2020comparative,medhat2020framework}

\begin{table}
 \renewcommand{\arraystretch}{1.35}
 \centering
 \caption{Summary of the Performance of ML-based TSP Techniques}
 \resizebox{1\textwidth}{!}{
 \begin{tabular}{llllrl}
  \hline
  \multirow{2}{*}{Study} & \multirow{2}{*}{Context} & \multirow{2}{*}{ML techniques} & \multirow{2}{*}{Feature Sets} & \multicolumn{2}{c}{~~~~~~~~~~~~~~~~Performance}\\
            &         &              &             & Metric & Value(s)\\
  \hline
  Busjaeger and Xie~\cite{busjaeger2016learning}       & TP  & SL \& NLP-based       & CVG + TXT + EXH     & Recall                   & [0\%-80\%]\\[-5pt] 
                                                      &     &                       &                     & APFD                     & [80\%-100\%]\\
  Lachmann et~al.~\cite{lachmann2016system}           & TP  & SL \& NLP-based       & CVG + TXT + EXH     & APFD                     & [52\%-92\%]\\
  Khalid et~al.~\cite{khalid2019weight}               & TP  & UL                    & INP                 & Accuracy                 & [79\%]\\
  Sharma et~al.~\cite{article}                        & TP  & SL                    & CPX                 & Precision                & [88\%-100\%]\\[-5pt]
                                                      &     &                       &                     & Recall                   & [80\%-100\%]\\[-5pt]
                                                      &     &                       &                     & F1-score                 & [89\%-100\%]\\
  Tonella et~al.~\cite{tonella2006using}              & TP  & SL                    & CVG + CPX + INP     & APFD                     & [88\%-94\%]\\
  Spieker et~al.~\cite{spieker2017reinforcement}      & TSP & RL                    & EXH                 & NAPFD                    & [0\%-100\%]\\
  
  Shi et~al.~\cite{shi2020reinforcement}              & TP & RL                    & EXH                 & NAPFD                    & [10\%-100\%]\\
  Rosenbauer et~al.~\cite{rosenbauer2020xcs}          & TP  & RL                    & EXH                 & NAPFD                    & [0\%-85\%]\\
  Bertolino et~al.~\cite{bertolinolearning}           & TSP  & RL \& SL              & CPX + EXH     & RPA                      &  [95\%-98\%] \\
  Jahan et~al.~\cite{jahan2019version}                & TP  & SL                    & CVG + CPX            & Accuracy                & [95.3\%-98.8\%]\\[-5pt]
                                                      &     &                       &                     & Precision                   & [94.5\%-98.9\%]\\[-5pt]
                                                      &     &                       &                     & Recall                 & [94.0\%-98.7\%]\\[-5pt]
                                                      &     &                       & 
                                            & APFD
                                        &
                              [61\%-97\%]\\
  Noor et~al.~\cite{noor2017studying}                 & TP  & SL                    & CVG + CPX + EXH     & Median First Fail Rank          & [19\%-30\%]\\
  Carlson et~al.~\cite{carlson2011clustering}         & TP  & UL                    & CVG + CPX + EXH     & APFD                     & [56\%-74\%]\\
  Singh et~al.~\cite{singhmachine}                    & TP  & SL                    & CPX                 & Quality Improvement      & [39\%] \\

  Thomas et~al.~\cite{thomas2014static}               & TP  & NLP-based             & TXT                 & APFD                     & [85\%-96\%]\\
  Aman et~al.~\cite{aman2020comparative}              & TP  & NLP-based             & TXT                 & APFD                     & [87.1\%-99.6\%]\\
  
  Chen et~al.~\cite{chen2018optimizing}               & TP  & SL                    & CVG + EXH           & APFDc                    & [56.4\%-99.9\%]\\
  Mahdieh et~al.~\cite{mahdieh2020incorporating}      & TP  & SL                    & CVG + CPX           & APFD                     & [46.9\%-71.0\%]\\
  Almaghairbe et~al.~\cite{almaghairbe2017separating} & TS  & UL                    & TXT                 & F1-score                 & [0\%-100\%]\\
  Lima et~al.~\cite{lima2020multi}                    & TP  & RL                    & EXH                 & NAPFD                    & [36.3\%,99.9\%]\\[-5pt]
                                                      &     &                       &                     & RMSE                     & [0.1\%,51.7\%]\\
  Lima et~al.~\cite{lima2020learning}                 & TP  & RL                    & EXH                 & NAPFD                    & [48.8\%-99.9\%]\\

  Chen et~al.~\cite{chen2011using}                    & TS  & UL                    & CVG                 & Difference of F1-score                 & [-20\%-40\%]\\
  Kandil et~al.~\cite{kandil2017cluster}              & TS  &UL \& NLP-based             & CVG + TXT                 & F1-score                 & [42\%-100\%]\\
  Hasnain et~al.~\cite{hasnain2019recurrent}          & TP  & SL                    & EXH                 & MAE                      & [0.17\%-8.23\%]\\
  Wang et~al.~\cite{wang2012using}                    & TS  & UL                    & CVG                 & Failure Detection Ratio  & [10\%-100\%]\\
  Palma et~al.~\cite{palma2018improvement}            & TP  & SL                    & CVG + CPX + EXH     & Median First Fail Rank          & [10\%-40\%]\\
  Yoo et~al.~\cite{yoo2009clustering}                 & TP  & UL                    & CVG + INP           & APFD                     & [14.4\%-99.6\%]\\
  Lima et~al.~\cite{do2020multi}                      & TP  & RL                    & EXH                 & NPAFD                    & [36.3\%,99.9\%]\\[-5pt]
                                                      &     &                       &                     & APFDc                    & [36.6\%-99.9\%]\\[-5pt]
                                                      &     &                       &                     & NTR                      & [0.3\%-73.3\%]\\
  Medhat et~al.~\cite{medhat2020framework}            & TSP & NLP-based             & CVG + TXT + EXH              & Precision                & [70\%-90\%]\\
  Mirarab et~al.~\cite{mirarab2008empirical}          & TP & SL                    & CVG + CPX           & APFD                     & [20\%-100\%] \\

  \hline
  \multicolumn{6}{l}{
     \begin{tabular}
            {@{}l@{}} 
                      {\footnotesize\textbf{Feature Sets:}
                      CVG=Coverage Information,
                      CPX=Code Complexity,
                      TXT=Textual Data,
                      EXH=Execution History,
                      INP=User Input}\\

                      {\footnotesize \textbf{ML Techniques:}
                                RL=Reinforcement Learning,
                                SL=Supervised Learning,
                                UL=Unsupervised Learning}\\
                                
                      {\footnotesize ~~~~~~~~~~~~~~~~~~~~~~~~
                                NLP-based=Natural Language Processing-based
                                }
     \end{tabular}
  }
 \end{tabular}
}
\label{tab:performance_all_papers}
\vspace{-10pt}
\end{table}

\begin{table}
\centering
\caption{Baselines Used to Evaluate ML-based TSP Techniques}
 \resizebox{1\textwidth}{!}{
 \begin{tabular}{p{10.3cm}l}
 \hline
 Baseline & Study \\
 \hline
 Random approach &
 \cite{spieker2017reinforcement,lima2020learning,busjaeger2016learning,jahan2019version,lachmann2016system,mahdieh2020incorporating,singhmachine,tonella2006using,thomas2014static} \\ 
 Traditional metric-based approach & \cite{carlson2011clustering,yoo2009clustering,noor2017studying} \\
 RETECS \cite{spieker2017reinforcement} & \cite{rosenbauer2020xcs,shi2020reinforcement} \\
 Genetic algorithm & \cite{lima2020multi} \\
 K-Means algorithm & \cite{chen2011using} \\
 LSTM and simple RNN & \cite{hasnain2019recurrent} \\
 Execution spectra-based sampling (ESBS) \cite{yan2010dynamic} & \cite{wang2012using} \\
 Cost-only technique (tests ordered based on execution time) & \cite{chen2018optimizing} \\
 Traditional and similarity-based metrics regression model \cite{noor2017studying} & \cite{palma2018improvement}\\
 Multi-Armed Bandit (MAB) \cite{robbins1952some} with random and greedy approaches & \cite{do2020multi} \\
 
 Not mentioned & \cite{bertolinolearning,almaghairbe2017separating,kandil2017cluster,khalid2019weight,mirarab2008empirical,article,aman2020comparative,medhat2020framework} \\
 \hline
 \end{tabular}
 }
\label{tab:baselines}
\end{table}

There are general differences between ML-based TSP techniques, including the online adaptation of RL techniques and the more straightforward training process of UL, since UL does not require labeling. However, given that ML-based TSP techniques have been evaluated using different datasets, it is not possible to precisely compare these techniques based on the reported performance results. To address this issue, we performed a statistical analysis to investigate which sets of features are associated with higher modeling performance for ML-based TSP techniques. To do this, we used the five groups of features used to train ML-based TSP techniques (described in RQ2), namely code complexity, textual data, coverage information, execution history, and user inputs.

In addition, due to the variety of metrics used to evaluate the performance of ML-based TSP techniques (described in RQ3), we limited our analysis to APFD and NAPFD as evaluation metrics. First, most (16 out of 29) of the studies used APFD and NAPFD as evaluation metrics, and studies based on other metrics are too few to draw statistically valid conclusions. Second, APFD and NAFPD measure TSP performance in a similar way, though APFD assumes all the faults are detected, whereas NAPFD considers the ratio between detected and total faults. Thus, when all test cases are executed and all faults are detected, NAPFD is equal to APFD. Lastly, as discussed in Section 3.3, general metrics, such as true positives, are defined and calculated differently across TSP studies, making comparisons of TSP performance using precision or recall meaningless. In contrast, primary studies defined APFD and NAPFD in a consistent way.
Further, we did not consider the studies by Thomas et~al.~\cite{thomas2014static} and Aman et~al.~\cite{aman2020comparative} in this analysis, since they only used NLP techniques for preprocessing textual data, and then prioritized test cases using non-ML algorithms. This resulted in considering 14 out of the 29 primary studies.

However, we further excluded four of those studies as they report NAPFD values using line plots~\cite{spieker2017reinforcement,rosenbauer2020xcs,shi2020reinforcement} or boxplots~\cite{mirarab2008empirical}, making it difficult to estimate the average performance of the techniques. As a result, we considered ten studies~\cite{do2020multi,lima2020learning,lima2020multi,carlson2011clustering,yoo2009clustering,busjaeger2016learning,jahan2019version,lachmann2016system,mahdieh2020incorporating,tonella2006using} in which APFD/NAPFD values were explicitly and precisely reported. Given that the performance of an ML-based TSP technique can differ from one subject (i.e., software project) to another, we extracted individual APFD or NAPFD values for each subject in all primary studies.
In total, we obtained 72 observations.
When a study used multiple experimental setups for a subject (e.g., different time budgets~\cite{do2020multi,lima2020learning,lima2020multi}, test suites~\cite{yoo2009clustering}, or reward functions~\cite{do2020multi,lima2020multi}), we computed the average NAPFD value of all setups.

While code complexity, coverage information, and execution history features are extensively used in the selected studies, we found that textual features were used in only two of them~\cite{busjaeger2016learning,lachmann2016system} (three subjects) and user input features were used once~\cite{tonella2006using} (one subject).
We used the APFD/NAPFD values as a dependent variable and the sets of features as independent variables.
Initially, we fitted a multivariable linear regression model to explain the NAPFD using the sets of features used as independent variables. However, due to the small sample size and strong associations between features and ML techniques, regression modeling results were difficult to interpret.
Therefore, we restricted ourselves to bivariate analysis to explore how significant is the effect of each feature set on NAPFD values and attempt to interpret the most plausible reasons for the results. In particular, we used the Mann-Whitney-Wilcoxon test~\cite{wilks2011statistical}, a non-parametric statistical test, with $\alpha = 0.05$, to test whether NAPFD values obtained with and without a feature set come from the same population. For each independent variable (feature set), we tested the null hypothesis that there is no significant difference between the distributions of NAPFD values obtained with and without this feature set. We used Cliff's delta effect-size measures~\cite{cliff1993dominance} to verify how practically significant is the difference in magnitude between two distributions of NAPFD values.

\begin{table}[hbt!]
\centering
\caption{Results of Wilcoxon Test and Effect Size (Cliff's Delta)}
\begin{tabular}{p{4.4cm}p{2.2cm}p{3cm}}
\hline
Feature Set & P-Value & Effect Size (\textit{delta})\\
 \hline
   Code Complexity    & \textbf{$<$ 0.0001}  & 0.734 (\textit{large})\\
   Coverage Information      & \textbf{$<$ 0.0001}  & 0.659 (\textit{large})\\ 
   Execution History  & \textbf{0.0008}  & -0.531 (\textit{large}) \\
   Textual Data       & 0.0806    & -\\ 
   User Input         & 0.5971     & -\\
 \hline
\end{tabular}
\label{tab:wilcoxon}
\end{table}

Table~\ref{tab:wilcoxon} presents the results of the Wilcoxon tests and effect sizes. Our statistical analysis reveals that code complexity, coverage information, and execution history have a statistically significant effect and large effect sizes on NAPFD. Probably due to the scarcity of data for these feature sets, we observe no significant relationship between the use of textual data or user input features and NAPFD. 
For example, user input is only used by eight subjects, whereas textual data is used by three subjects. All these subjects also involve coverage information features besides user inputs. In addition, seven of those subjects employed UL and one of them employed SL.
For all subjects that use textual features, SL is the sole ML technique employed and is always combined with coverage information and execution history features.
Therefore, future work should experiment other types of ML techniques (e.g., RL) using textual data or user input features in order to investigate their impact on NAPFD.

\begin{figure}
     \centering
     \begin{subfigure}[b]{0.3\textwidth}
         \centering
         \includegraphics[width=\textwidth]{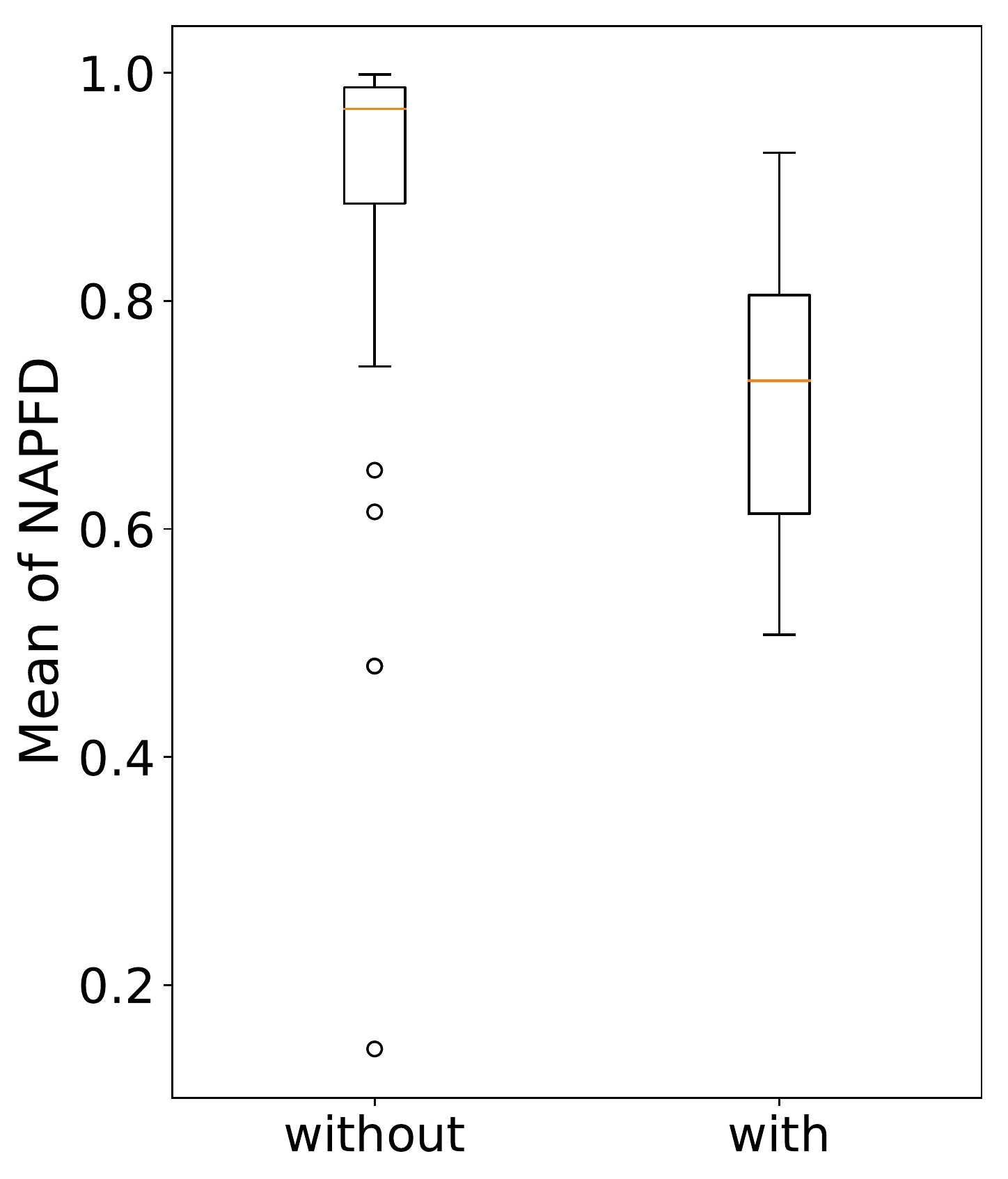}
         \caption{Code Complexity Features}
         \label{fig:code_complexity}
     \end{subfigure}
     \hfill
    \begin{subfigure}[b]{0.3\textwidth}
         \centering
         \includegraphics[width=\textwidth]{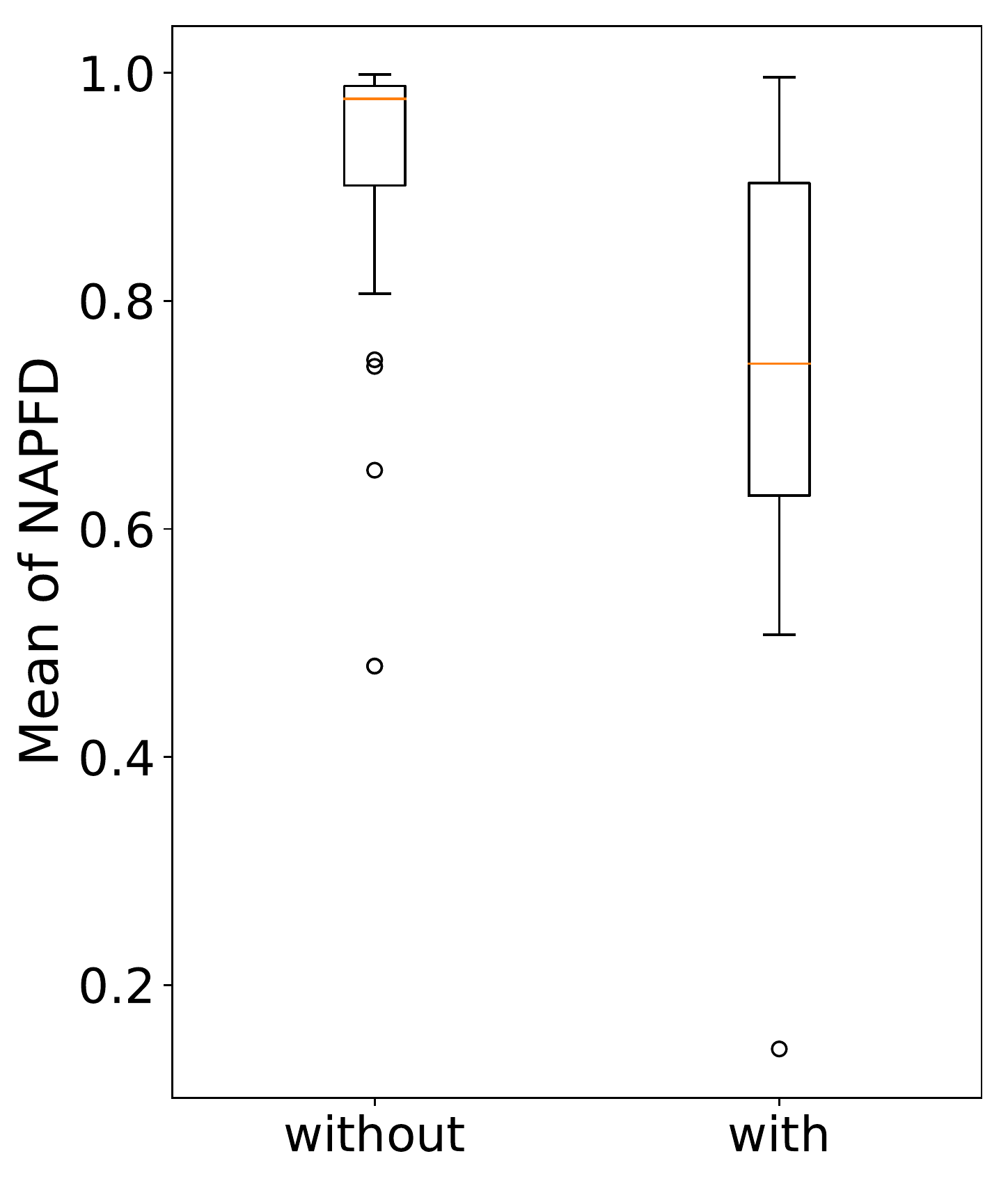} 
         \caption{Coverage Features}
         \label{fig:code_coverage}
     \end{subfigure}
     \hfill
    \begin{subfigure}[b]{0.3\textwidth}
        \centering
        \includegraphics[width=\textwidth]{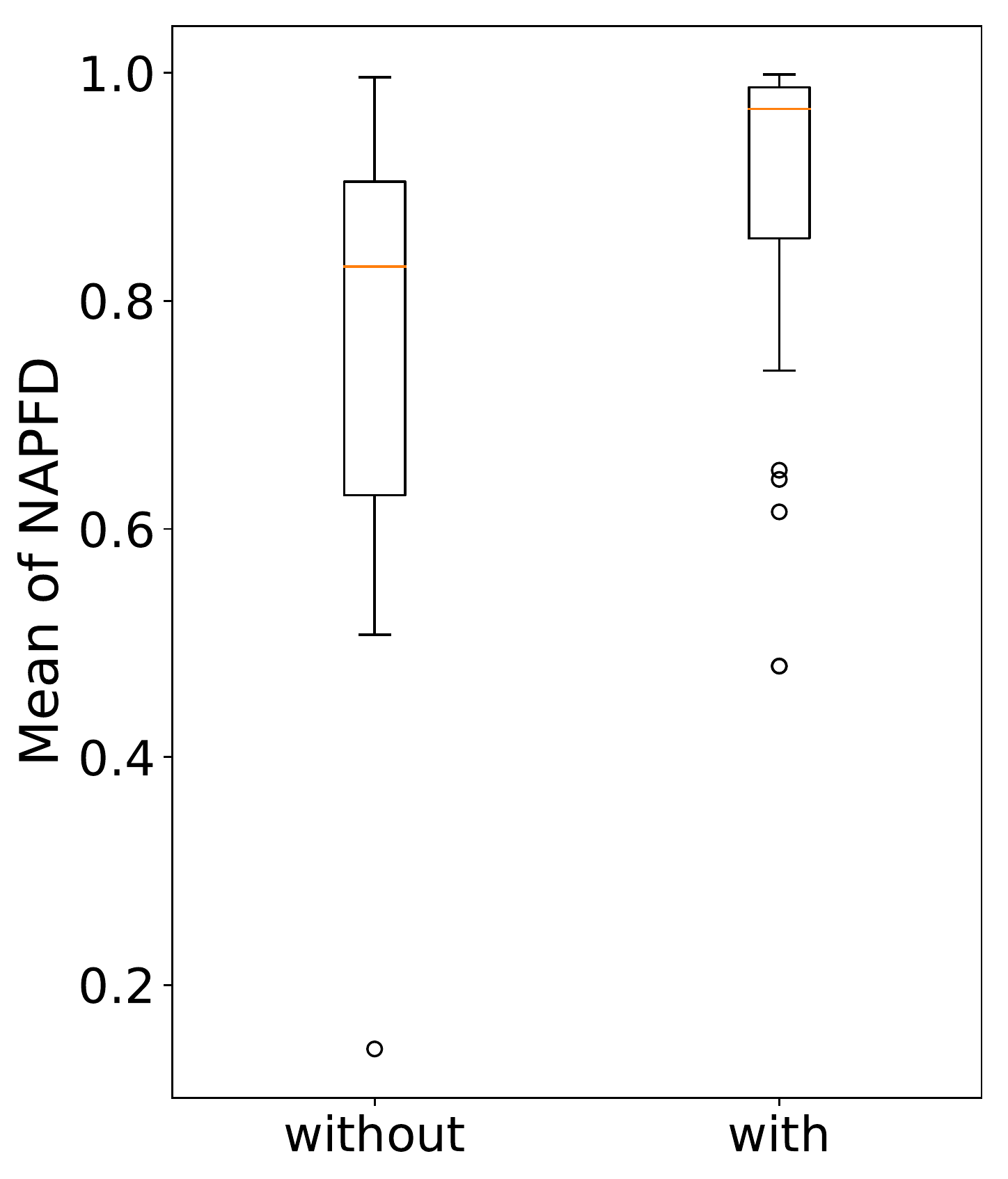}
        \caption{Execution History Features}
        \label{fig:execution_history}
     \end{subfigure}
     \hfill
     \caption{Boxplots for Code Complexity, Coverage Information, and Execution History Feature Sets}
    \label{fig:boxplots}
    \vspace{-10pt}
\end{figure}

Figure~\ref{fig:boxplots} provides the boxplots for code complexity, coverage information and execution history. The median NAPFD values, with and without execution history, are 0.968 and 0.830, respectively. This suggests that using execution history features significantly increases the likelihood of producing higher NAPFD values.
However, looking at the median NAPFD values with (0.730) and without code complexity (0.968), we observe that the former is unexpectedly associated with lower NAPFD values. Similarly, the median NAPFD values with and without coverage information are 0.745 and 0.977, respectively.
To investigate the reasons behind such surprising results, we performed additional analyses of the different combinations of feature sets and ML techniques in order to determine if there are potential confounding factors.

\begin{table}[hbt!]
\centering
\caption{Number of Observations and Median Test Suite Sizes of Feature Set Combinations. UL: Unsupervised Learning, SL: Supervised Learning, RL: Reinforcement Learning.}
\resizebox{1\linewidth}{!}{\begin{tabular}{llrr}
\hline
Feature Set Combination &ML Technique &\# of Subjects  & Median Test Suite Size \\
 \hline
   \multirow{3}{*}{Code Complexity}                                       & UL & 2 &  \multirow{3}{*}{656} \\
                                                                          & SL & 11 &  \\
                                                                          & RL & 0 &  \\[4pt]
   \multirow{3}{*}{Coverage Information}                                  & UL & 9 &  \multirow{3}{*}{523} \\
                                                                          & SL & 14&  \\
                                                                          & RL & 0 &  \\[4pt]
   \multirow{3}{*}{Execution History}                                     & UL & 2 &  \multirow{3}{*}{57} \\
                                                                          & SL & 3 &  \\
                                                                          & RL & 49&  \\[4pt]
   \multirow{3}{*}{Code Complexity \& Coverage Information}               & UL & 2 &  \multirow{3}{*}{656} \\
                                                                          & SL & 11&  \\
                                                                          & RL & 0 &  \\[4pt]
   \multirow{3}{*}{Code Complexity \& Execution History}                  & UL & 2 &  \multirow{3}{*}{782} \\
                                                                          & SL & 0 &  \\
                                                                          & RL & 0 &  \\[4pt]
   \multirow{3}{*}{Coverage Information \& Execution History}             & UL & 2 &  \multirow{3}{*}{656} \\
                                                                          & SL & 3 &  \\
                                                                          & RL & 0 &  \\[4pt]
   \multirow{3}{*}{Code Complexity \& Coverage Information \& Execution History} & UL & 2 &  \multirow{3}{*}{782} \\
                                                                          & SL & 0 &  \\
                                                                          & RL & 0 &  \\
   \hline
\end{tabular}
}
\label{tab:comparison}
\vspace{-8pt}
\end{table}

Table~\ref{tab:comparison} presents different combinations of code complexity, coverage information, and execution history, and ML techniques, along with their corresponding number of observations and the median of test suite sizes. We observed, as presented above, that using both code complexity and coverage information features is significantly associated with lower NAPFD values. However, when we investigated the subjects that use either or both these feature sets, we found that none of the subjects employ RL models, but rather UL and SL models.

In contrast, we found that the subjects that use execution history features have significantly higher NAPFD values. Most of these subjects (91\%) employ RL models, which were all used in recent TSP studies (e.g., six out of seven papers were published in 2020). Due to this high degree of association, it is hard to say whether higher performance is due to one or both of these factors. For example,  we cannot conclude that the reasons for lower NAPFD values are inadequate ML techniques or insufficient features. Therefore, the performance of ML techniques and feature sets need to be investigated independently, without confounding effects, which is not possible with the currently available data.

Our analysis also shows that only a few ($<$ 7\%) subjects combine coverage information and complexity features with execution history features, which indicates that subjects in primary studies tend to contain partial feature sets; yet, none of those subjects employ RL models.

As a result, future research should apply a variety of learning techniques (including RL) based on a comprehensive set of features (including coverage information and complexity, execution history) on a large number of subjects. This would render possible the investigation, without confounding effects, of the combined impact of feature sets and learning techniques on TSP performance.

To further investigate the possible reasons behind the association of execution history features with higher NAPFD values,
we analyzed the test suite size of the subjects that use and do not use these features. We observed that the former have significantly smaller test suite sizes than the other subjects ($p-value<0.0001$ and large effect size: $delta$ = 0.613). For those subjects that use execution history features, we observe no significant relationship between their test suite size and the obtained NAPFD values ($p-value$ = 0.521). Still, we should note that evaluating TSP techniques on a small sample size may not be sufficient to demonstrate their practical advantage. This result suggests that future research should consider developing and evaluating TSP techniques using execution history features on larger datasets, to investigate whether such features would lead to higher NAPFD values. Moreover, considering that test failure rates can impact NAPFD values, we analyzed the failure rate data of 49 subjects from three studies~\cite{do2020multi,lima2020learning,lima2020multi} that employ RL models using execution history features. We observed that the higher the failure rate, the lower the NAPFD value ($p-value<0.0001$). One explanation is that, though questionable, it is common practice to report APFD/NAPFD values of $one$ when no test case fails, as discussed in Section~\ref{RQ3}. Nevertheless, this result suggests that there is still ample room for developing more accurate TSP techniques that can handle tests with higher failure rates. Also, studies should report failure rates alongside NAPFD values to enable proper interpretation of results.

\mybox{\textbf{\textit{RQ4 Conclusion:}} Comparing and interpreting the performance of ML-based TSP techniques is challenging because (a) there is no agreed-upon, consistently used performance metric, (b) test suite sizes and failure rates widely vary across studies, and (c) there are confounding factors. Comprehensive studies must be designed and run to prevent confounding factors and investigate all the interaction effects between feature sets, ML techniques, failure rates, and test suite sizes, as well as their impact on performance.}

\subsection{\textbf{RQ5.} Are ML-based TSP studies repeatable and reproducible?} 
In the context of ML-based TSP, we label a paper as \textit{repeatable} if we can rerun the exact same experiments, using the same datasets, based on the information in the paper. We also label a paper as \textit{reproducible} if it is repeatable and we can check whether the results obtained and the conclusions drawn are statistically equivalent to the ones reported in the paper. In the context of ML-based TSP, we define an experiment as being composed of four steps: (1) training an ML model based on data from previous builds (i.e., versions), (2) applying the trained model on the next builds to select/prioritize test cases, (3) running the test cases based on the selection/prioritization output, and (4) analyzing the results based on the defined evaluation metrics (e.g., APFDs). 
In this question, we aim to analyze the extent to which data, artifacts, and instructions shared by the papers enable the validation and confirmation of reported results and conclusions. Thus, we examine whether or not the papers provide the following artifacts and information. 
 
\begin{itemize}
\label{sec:M1-M7}
    \item \textit{\textbf{M1.} Training and evaluation datasets.} These datasets are the most basic requirement. Thus, we mark all papers as not repeatable if they do not provide them (e.g., as CSV or JSON files), without any further investigation.  
    
    \item \textit{\textbf{M2.} Scripts to train and test the model.} These scripts enable researchers to train and evaluate the ML models that are the core of any ML-based TSP work. The papers also need to provide instructions on how to execute these scripts. 
    
    \item \textit{\textbf{M3.} Details about ML algorithms and relevant hyper-parameters.} The papers need to report the used ML algorithms, libraries, and tools along with their versions, hyper-parameter values, and the required execution environment (e.g., OS and memory requirements). If the used library provides more than one method for a certain task, the function name and related parameters should be reported as well. Such information enables the training and evaluation of ML models if the relevant scripts are not provided.  
    
    \item \textit{\textbf{M4.} A detailed description of the experiments.} This enables researchers to run the experiments in the same way as the paper in terms of experimental setup (e.g., time budgets or reward functions), validation method (e.g., cross-validation~\cite{stone1974cross}), and the method of controlled random seeds when it is used. Further, almost all ML algorithms entail an inherent degree of randomness~\cite{liem2020run}. Thus, the papers should report how such randomness was addressed, for example by repeating experiments multiple times, without which achieving reproducibility is not possible.
    
    \item \textit{\textbf{M5.} Scripts to automate the experiments.} The paper should provide scripts that automate the experiments. This greatly simplifies their repetition and prevents mistakes due to the re-implementation of the scripts and infrastructure. 
    
    \item \textit{\textbf{M6.} Reporting the complete evaluation results per subject.} To compare the results after rerunning the experiment of a paper, it is important to provide the evaluation results (e.g., the average performance of multiple runs) for each experiment. The studies that provided evaluation results using only plots do no satisfy M6.
    
    \item \textit{\textbf{M7.} Specifying how conclusions are drawn.} In addition to reporting results per subject based on defined evaluation metrics, papers often draw conclusions concerning the practicality and effectiveness of the proposed approaches in comparison with each alternative and baseline. Thus, the paper should report how such conclusions are made by describing the used baselines, applied statistical analysis techniques (e.g., statistical significance test and effect size measurement), and the number of runs for each experiment.
\end{itemize}

To summarize, we consider a paper to be repeatable if it provides M1, M2, or M3, and M4 or M5. It is reproducible if it provides M1, M2, or M3, M4 or M5, M6, and M7. Otherwise, it is not repeatable. Note that we do not rerun the experiments and our analysis is only based on assessing the degree of support for M1-M7. Also, we ignore two strict conditions for improving repeatability and reproducibility, respectively, since most papers do not meet them, e.g., reporting random seeds and providing detailed results for each experiment. 

\begin{table}
\centering
\caption{Artifacts and Information Provided by the Papers. M2: Training and testing scripts , M3: Hyperparameters and libraries, M4: Experiment description, M5: Experiment scripts, M6: Complete evaluation results, M7: Statistical analysis (see Section~\ref{sec:M1-M7} for more details).}
\label{tab:reproduciblity}
\begin{tabular}{p{3.20cm}p{0.5cm}p{0.5cm}p{0.5cm}p{0.5cm}p{0.5cm}p{0.5cm}p{2.85cm}}

 \hline
  & M2 & M3 & M4 & M5 & M6 & M7 & Label \\
 \hline
 Bertolino et~al.~\cite{bertolinolearning}            & Yes & No & Yes & Yes & Yes & Yes & Reproducible\\
 Palma et~al.~\cite{palma2018improvement}             & Yes & No & Yes & Yes & Yes & Yes & Reproducible\\
 Mahdieh et~al.~\cite{mahdieh2020incorporating}       & Yes & No  & No & Yes & Yes & Yes & Reproducible\\
 Aman et~al.~\cite{aman2020comparative}               & Yes & No  & Yes  & No  & Yes & Yes & Reproducible\\
 Thomas et~al.~\cite{thomas2014static}                & Yes  & No  & Yes  & Yes  & Yes & Yes & Reproducible\\
 
 Rosenbauer et~al.~\cite{rosenbauer2020xcs}           & Yes & Yes & Yes & Yes & No  & Yes & Repeatable\\
 Spieker et~al.~\cite{spieker2017reinforcement}       & Yes & No  & Yes & Yes & No  & No  & Repeatable \\
 
 Lima et~al.~\cite{lima2020multi}                     & No & No & Yes & No & Yes & Yes & Not Repeatable\\
 Lima et~al.~\cite{lima2020learning}                  & No & No & Yes & No & Yes & Yes & Not Repeatable\\
 Lima et~al.~\cite{do2020multi}                       & No  & No & Yes & No  & Yes & Yes & Not Repeatable\\
 Shi et~al.~\cite{shi2020reinforcement}               & No  & No & Yes & No  & No  & No  & Not Repeatable \\
 Noor et~al.~\cite{noor2017studying}                  & No  & No  & Yes  & No  & No & Yes & Not Repeatable\\
  \hline

 \hline
\end{tabular}
\label{tab:R&R2}
\end{table}

\begin{figure}
    \centering
    \includegraphics[width=0.75\columnwidth]{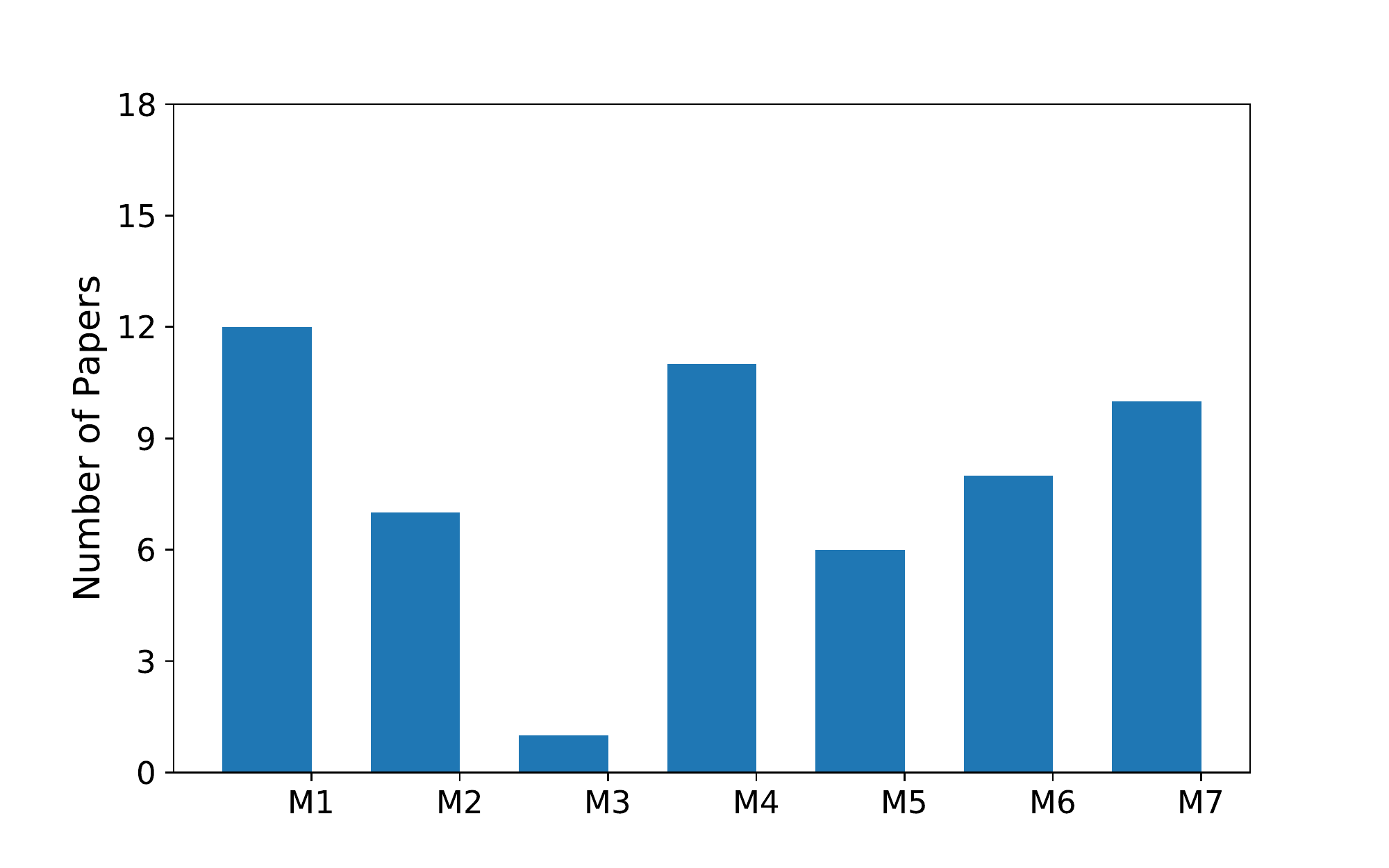}
    \vspace{-7pt}
    \caption{Information Provided by Papers}
    \label{fig:rprd}
\end{figure}

As listed in Table~\ref{tab:reproduciblity}, 12 of the papers made their datasets available (they satisfy M1). We examined these papers to assess whether they meet criteria M2-M7. Table~\ref{tab:reproduciblity} shows detailed results regarding M2-M7 for each paper, and Figure~\ref{fig:rprd} summarizes M1-M7 results for all these 12 papers. The results show that only five out of the 12 papers (42\%) are reproducible.
This is far from ideal, since most ML-based TSP papers reported empirical results that cannot be confidently validated or reproduced. In addition, only two papers are repeatable, thus enabling the community to confirm the reported conclusions by running the exact same experiments. The remaining five papers, though they provide their datasets, are not repeatable as they miss technical details (e.g., language or library versions), thus preventing them from meeting important repeatability criteria.
Moreover, 21 of the papers used subjects from publicly available datasets, such as the Software-artifact Infrastructure Repository (SIR).\footnote{\url{https://sir.csc.ncsu.edu}} Such datasets include tests cases that were collected from code changes made years ago (before 2016), which do not reflect current practices in continuous development and integration. Hence, future research should collect data on more recent subjects from a CI context (e.g., extending TravisTorrent~\cite{msr17challenge}) with a more comprehensive set of features, higher failure rates, and larger execution times.

Based on the above analysis, we argue that the situation regarding repeatability and reproducibility is alarming as it undermines our capacity, as a scientific community, to grow a reliable and interpretable body of knowledge. This issue needs to be addressed by providing detailed guidelines and instructions to be used consistently across papers in order to make experiments reproducible.
Mattis et~al.~\cite{mattis2020rtptorrent} surveyed all datasets used by prior studies in the context of test case prioritization. Besides, the authors proposed RTPTorrent, a dataset for evaluating test prioritization based on TravisTorrent. Yet, though recently published, RTPTorrent is limited to 20 Java projects and contains no features related to coverage information, user input, and textual data. In addition, data in RTPTorrent does not reflect current CI practices, since it was collected from code changes committed in 2011-2016.
Therefore, future work should construct up-to-date benchmarks with more comprehensive feature sets for a thorough evaluation of ML-based TSP techniques.

\mybox{\textbf{\textit{RQ5 Conclusion:}} Only a small number of ML-based TSP studies are reproducible. Further, the currently available datasets used by ML-based TSP techniques contain varying feature sets based on data collected from code changes made before 2016, which may not represent current CI practices. Hence, reproducible studies with more appropriate datasets are needed to better assess the performance of TSP techniques and develop a usable body of knowledge regarding TSP over time.}

\subsection{ML-based TSP Summary} 
\begin{figure}[ht!]
    \centering
    \vspace{-30pt}
    \includegraphics[width=1\textwidth]{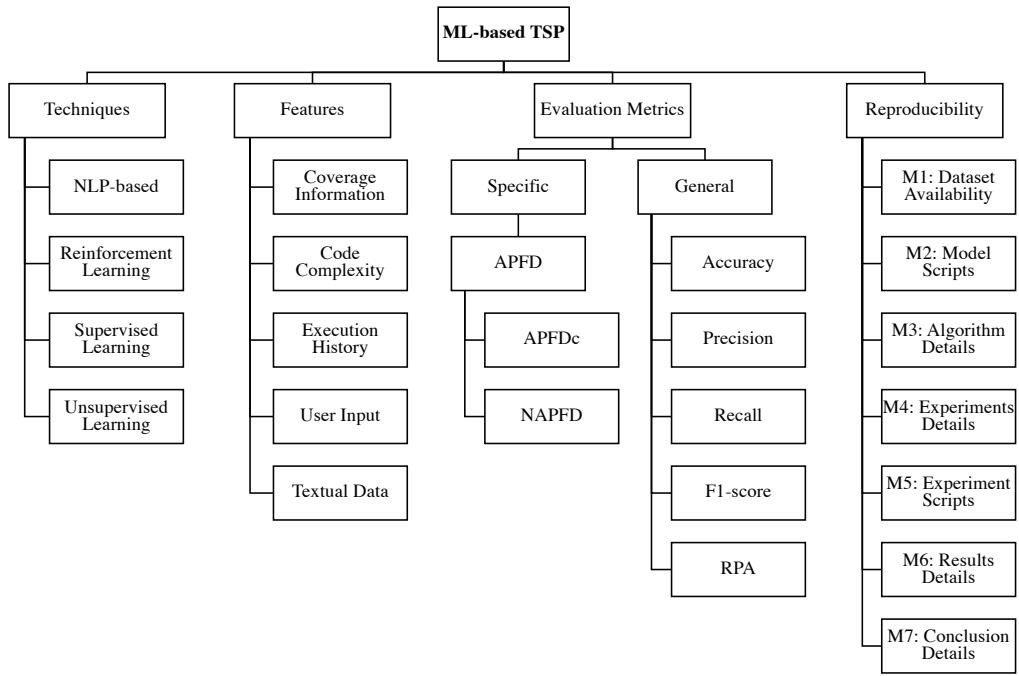}
    \vspace{-40pt}
    \caption{Summary of ML-based TSP}
    \label{fig:taxonomy}
\end{figure}

We summarize the results related to our research questions at a high-level in Figure~\ref{fig:taxonomy}. This summary characterizes existing TSP techniques and the empirical methods that were followed to evaluate them.
In particular, the summary shows the ML techniques used for TSP, the feature sets used to train models, the metrics used to evaluate the results of the models, and the reproducibility criteria used to consider a study as repeatable or reproducible.
This summary can be used as a taxonomy for classifying future TSP studies.

\section{Discussion}
\label{sec:discussion}
This section discusses the implications of our findings based on primary studies.

\vspace{3pt}
\noindent\textbf{Subjects, Feature Sets, or ML Models.}
Our analysis shows that the performance of ML-based TSP techniques depends on several factors, such as the subjects (software projects) used for evaluation, the ML models used to select/prioritize test cases, and the features used to train the models. Subjects have different characteristics, such as test suite size, history, and failure rate. Also, a set of features may work well on one project but not on another project, thus suggesting that combining all sets of features may be beneficial. ML models may differ in terms of being supervised or unsupervised, which might make them only applicable to certain types of datasets (e.g., labeled and unlabeled).
Therefore, to improve TSP results, future research should take into consideration (a) collecting data from subjects that have considerable numbers of test cases with varying but plausible failure rates and regression testing time, (b) computing as many relevant features as possible about test cases and their covered code, and (c) investigating the application of advanced NLP-based similarity analysis techniques (e.g., BERT~\cite{devlin2018bert} and CodeBERT~\cite{feng2020codebert}) for more effective TSP.

\vspace{5pt}
\noindent\textbf{CI as a Source and Target for TSP.} 
CI increases the frequency of running software builds, including regression tests. CI produces rich historical data about test case executions under various runtime environments. According to our statistical analysis, execution history features are significantly associated with higher TSP performance. Therefore, training ML models using CI-generated data is likely to improve the performance of TSP techniques.
In practice, the high frequency of running CI builds warrants the use of TSP models. However, little is known about the practical challenges of adopting TSP techniques. Despite the lack of clear and reliable empirical results, researchers should increase the awareness of developers about current state-of-the-art techniques for TSP. Doing so does not only helps developers to leverage the benefits of TSP techniques in their CI pipelines, but also helps researchers receive feedback and gain insights from developers about the practical effectiveness and possible improvements of TSP techniques.

\vspace{5pt}
\noindent\textbf{Standards and Benchmarks for Evaluating TSP.} Our analysis reveals no standard practices regarding the design of experiments, features, and evaluation metrics used to assess the performance of TSP techniques. TSP evaluation metrics can be computed differently, making it difficult to compare results across studies. Only a few studies build on previous studies in designing their experiments and evaluating their techniques (e.g.,~\cite{rosenbauer2020xcs} and~\cite{spieker2017reinforcement}). Comparing TSP techniques relying on published papers is therefore difficult. Ideally, one should reproduce the results using the same datasets and experimental setup. However, our analysis shows that only a few papers are reproducible. Therefore, future research should pay more attention to constructing adequate benchmarks to evaluate TSP techniques. Moreover, researchers should develop standards to ensure the trustworthiness of the results reported by TSP studies. For example, subjects should comply with a minimal failure rate and regression testing time in order to be suitable for evaluating TSP techniques.

\vspace{5pt}
\noindent\textbf{Application Scenarios of ML-based TSP Techniques.}
In the context of CI, RL-based TSP techniques are more suitable to apply than UL-based and SL-based techniques, since they integrate new data into already constructed models without retraining them from scratch.
The majority of SL-based techniques are restricted to the classical batch setting and assume full datasets to be available before training, which does not enable incremental learning (i.e., integrating new data into already constructed models) but rather reconstruct new models from scratch. This is not only very time-consuming but also leads to potentially outdated models, which can be problematic, especially in a CI context.
Clustering algorithms (UL techniques) require expensive tuning, regarding distance metrics and the number of clusters, which affects their effectiveness in a CI context. Also, the computational complexity of UL techniques is high as, for most of them, finding the optimal solution is an NP-hard problem, which can cause scalability issues.
Finally, despite the potential of NLP-based techniques, their application in a TSP context is very limited. NLP techniques should be utilized to devise effective similarity-based TSP techniques, since coverage analysis in a CI context can be computationally expensive. Many textual software development artifacts are a rich source of information to devise useful features for training more effective NLP-based models for TSP.

\section{Related Work}
\label{sec:relatedwork}
We identified two Systematic Literature Reviews (SLRs), two systematic mapping studies, and one review, which are related to TSP using machine learning techniques, as listed in Table~\ref{tab:relatedreviews}. In Table~\ref{tab:relatedreviews}, we show the focus of each study, publication year, covered years, and the number of included primary studies.

Kazmi et~al.~\cite{kazmi2017effective} focused on TS techniques. They reviewed 47 empirical studies covering the period from 2007 to 2015 and categorized them based on regression TS techniques that focused on cost, coverage, or fault-based effectiveness. 
The results of their work showed that multi-objective criteria (combination of cost-based, coverage-based, and fault-based criteria), are used for the evaluation of ML-based TS techniques.

Durelli et~al.~\cite{durelli2019machine} conducted a systematic mapping study on ML techniques applied to software testing based on 48 primary studies covering the period from 1995 to 2018. They classified them according to study type (publication type and research type), testing activity (compatibility testing, test case design, TP, etc.) and the types of ML techniques used. They concluded that ML techniques were mainly applied for test case generation, refinement, evaluation, and predicting the cost of testing-related activities.

The studies of Lima et~al.~\cite{lima2020test}, Khatibsyarbini et~al.~\cite{khatibsyarbini2018test} and Shaheamlung et~al.~\cite{shaheamlung2020comprehensive} all focused on TP techniques. Lima et~al.~\cite{lima2020test} reviewed 35 primary studies covering the period from 2009 to 2019. They presented the main characteristics of TP in CI environments (TCPCI) by classifying those studies according to the goal and type of TP techniques and the evaluation measures. They also accounted for CI-specific challenges and testing problems. Their results indicated a growing interest in TCPCI research and that 80\% of the techniques were history-based. They also observed a rising trend in the application of Artificial Intelligence (AI) techniques like ML for TCPCI. Eight studies used search-based or ML techniques. 
Khatibsyarbini et~al.~\cite{khatibsyarbini2018test} investigated and classified TP approaches based on 69 studies covering the period from 1999 to 2016. They concluded that TP approaches still had significant room for improvement.  
Shaheamlung et~al.~\cite{shaheamlung2020comprehensive} analyzed different prioritization techniques based on 22 studies covering the period from 2010 to 2019. They characterized them according to the type of techniques, their performance, and how they can be used in future research. They concluded that ML was being widely used for TP and showed better outcomes than other approaches.

In comparison with the studies mentioned above, our systematic review focuses exclusively on the application of ML techniques to the TSP problem. We performed a more thorough analysis of the usage and performance of ML techniques, combined with different feature sets, based on which we provide insights regarding the achievements, advantages, and current limitations of the application of ML for TSP. We also outline useful research directions. 

\begin{table}[ht]
\centering
\caption{Review Papers on TSP}
\resizebox{1\textwidth}{!}{
\begin{tabular}{llp{4.15cm}p{2cm}p{1.5cm}p{1.5cm}}
 \hline
 Reference & Study Type & Study Focus & Publication Year & Years Covered & Included Papers\\
 \hline
 Kazmi et~al.~\cite{kazmi2017effective}                   & Systematic literature review & Effective regression TS                      & 2017 & 2007-2015 & 47\\
 Khatibsyarbini et~al.~\cite{khatibsyarbini2018test}      & Systematic literature review & TP approaches in regression testing          & 2018 & 1999-2016 & 69\\
 Vinicius H. et~al.~\cite{durelli2019machine}             & Systematic mapping study     & Machine Learning applied to software testing & 2019 & 1995-2018 & 48\\
 Lima et~al.~\cite{lima2020test}                          & Systematic mapping study     & TP in continuous integration environments    & 2020 & 2009-2019 & 35\\
 Shaheamlung et~al.~\cite{shaheamlung2020comprehensive}   & Review         & TP in software engineering                                 & 2020 & 2010-2018 & 22\\
 \hline
\end{tabular}
}
\label{tab:relatedreviews}
\end{table}

\section{Threats to Validity}
\label{sec:threats}
The process used in this work is composed of multiple steps, including the search strategy, study selection, and data synthesis/extraction. While some parts of the process were automated (e.g., search queries and Covidence screening), other parts (e.g., data synthesis/extraction) were manual.
As a result, our search might have missed some relevant studies or information as our review covers a vast array of possible techniques and application contexts. To mitigate this issue, we considered several measures, as follows. 

\subsection{Internal Validity}
    \noindent\textbf{Search strings.} We used general terms related to machine learning or test case selection and prioritization in our search queries. Then, we selected the papers by applying inclusion and exclusion criteria that required reading at least the title and abstract of the papers. Still, our search terms might not be comprehensive enough. 
    To mitigate this issue, we also used more specific ML-related terms, such as `regression', `neural network', and `reinforcement learning'.\\

    \noindent\textbf{Study inclusion.} Including or excluding a study can be subjective. To mitigate this issue, we used the Covidence tool to screen the collected papers and exclude any duplicates. The inclusion/exclusion of papers was performed collaboratively by two co-authors together using well-defined criteria to filter out irrelevant studies.\\ 
    
    \noindent\textbf{Data extraction.} Manually extracting relevant information from papers may suffer from missing or misinterpreting information about the studies. To mitigate this issue, one co-author performed the data extraction from all papers. A second co-author validated the data extraction for papers when information was uncertain or hard to interpret. We resolved any confusion in the data extraction in discussion meetings with the other co-authors. 

\subsection{External Validity}
    \textbf{Literature repositories.} Our search result is limited to the online repositories used as data sources. To mitigate this issue, we used online repositories that have been extensively used by previous survey papers in software engineering and include well-known and leading software engineering venues.

\subsection{Conclusion Validity}
    \noindent\textbf{Paper conclusions.} The primary studies used different subjects and a variety of evaluation metrics. This made it difficult to identify clear patterns and draw definitive conclusions. To mitigate this issue, we performed a statistical analysis in which we assess the extent to which features, subjects, and ML models are associated with good TSP performance.

\subsection{Reliability Validity}
    \noindent\textbf{Study replication.} Replicating the study presented in this paper is possible as long as all steps of the search process presented in Section~\ref{sec:reserachmethod} and shown in Figure~\ref{fig:steps1} are followed. However, there could be some inconsistencies in the obtained results due to potential differences in opinions regarding data extraction. To mitigate this issue, we described every step in our search process in detail, provided a high-level summary of ML-based TSP that can be used as a taxonomy for classifying future TSP studies, and made publicly available a replication package~\cite{replicationpackage} containing the search strings and the data extracted from primary studies.

\section{Conclusion}
\label{sec:conlcusion}
In this paper, we present the results of a systematic literature review (SLR) regarding the application of Machine Learning (ML) techniques for Test case Selection and Prioritization (TSP). This is a particularly important topic in Continuous Integration (CI) contexts, with frequent builds and regression testing campaigns, and ML enables the optimal combination of various sources of information to effectively address the TSP problem. 
The review was conducted by following standard steps, including the definition of research questions, a search strategy, study selection criteria, and data synthesis and extraction method.

The search returned 29 primary studies covering the period from 2006 to 2020, plus a number of systematic reviews and mapping studies. We observed an increasing number of studies in recent years, which indicates a growing interest in this research area.

The main goal of this paper was to analyze how ML techniques have been used and assess what they have achieved and their limitations. We investigated the type of ML techniques used for TSP, hyper-parameters configuring ML algorithms, the experiment design and evaluation metrics of each study, and their experimental results. Then, we extracted and synthesized the data from the studies to address five research questions.

\vspace{5pt}
From our systematic review, we can conclude the following: 

\begin{enumerate}
    \item [a)] A variety of ML techniques were used for TSP: Supervised Learning, Unsupervised Learning, Reinforcement Learning, and NLP-based are the four main ones. Some combinations of ML techniques were also reported. For example, NLP-based techniques, which are often used for feature pre-processing, were combined with supervised or unsupervised learning to achieve better performance for test case prioritization. 
    More advanced ML techniques should further be investigated, such as the use of advanced NLP-based techniques (e.g., BERT~\cite{devlin2018bert} and CodeBERT~\cite{feng2020codebert}) to devise more effective similarity-based TSP techniques.

    \item [b)] Different types of features were used throughout the studies, including code complexity, execution history, coverage information, user inputs, and textual data. However, some test cases were designed for special contexts (e.g., web application) and led to domain-specific features (e.g., number of tokens, number of hyperlinks in a web page). Using such features, combined with ML techniques, can also lead to good results. We found that most of the TSP techniques used only historical data. Coverage information and complexity features were also occasionally and partially used, relying only on a few complexity metrics and similarity-based coverage. Thus, a comprehensive analysis of features and their impact on TSP accuracy appears to be a necessary and useful future research endeavor.

    \item [c)] There has been a growing interest in the application of RL to TSP in recent years. The focus of these investigations has been on the impact of different reward functions and policy learning techniques on RL performance. The main reason for the increasing use of RL is practical, that is its capacity to seamlessly handle change in the system and test suites. We also found that existing RL work mainly used execution history features, and therefore using a more extensive feature set could be particularly beneficial in that context. 
\end{enumerate}

Finally, the primary studies used different subjects---many of them questionable given the objectives of TSP---and different evaluation metrics. The lack of standard evaluation procedures and appropriate publicly available subjects makes it very challenging to draw reliable conclusions concerning ML-based TSP performance, which is necessary to characterize the state-of-the-art and identify open problems. Thus, getting the research community to converge towards common evaluation procedures, metrics, and benchmarks is vital for building a strong body of knowledge we can rely on, without which advancing the state-of-the-art remains an elusive goal.

\section*{Acknowledgement}
This work was supported by a research grant from Huawei Technologies Canada Co., Ltd, as well as by the Mitacs Accelerate Program, the Canada Research Chair and Discovery Grant programs of the Natural Sciences and Engineering Research Council of Canada (NSERC).

\bibliographystyle{unsrt}
\bibliography{refrences}

\begin{thebibliography}{10}

\bibitem{fowler2006continuous}
Martin Fowler and Matthew Foemmel.
\newblock {Continuous integration}.
\newblock {\em
  \url{http://www.dccia.ua.es/dccia/inf/asignaturas/MADS/2013-14/lecturas/10\_Fowler\_Continuous\_Integration.pdf}},
  2006.

\bibitem{khatibsyarbini2018test}
Muhammad Khatibsyarbini, Mohd~Adham Isa, Dayang~NA Jawawi, and Rooster Tumeng.
\newblock Test case prioritization approaches in regression testing: A
  systematic literature review.
\newblock {\em Information and Software Technology}, 93:74--93, 2018.

\bibitem{lima2020test}
Jackson A~Prado Lima and Silvia~R Vergilio.
\newblock Test case prioritization in continuous integration environments: A
  systematic mapping study.
\newblock {\em Information and Software Technology}, 121:106268, 2020.

\bibitem{ghaleb2019empirical}
Taher~A Ghaleb, Daniel~A Da Costa, and Ying Zou.
\newblock An empirical study of the long duration of continuous integration builds.
\newblock In {\em Empirical Software Engineering}, 24(4): 2102--2139, 2019.

\bibitem{rothermel2001prioritizing}
Gregg Rothermel, Roland~H. Untch, Chengyun Chu, and Mary~Jean Harrold.
\newblock Prioritizing test cases for regression testing.
\newblock {\em IEEE Transactions on software engineering}, 27(10):929--948,
  2001.

\bibitem{kim2002history}
Jung-Min Kim and Adam Porter.
\newblock A history-based test prioritization technique for regression testing
  in resource constrained environments.
\newblock In {\em Proceedings of the 24th international conference on software
  engineering}, pages 119--129, 2002.

\bibitem{kitchenham2004procedures}
Barbara Kitchenham.
\newblock Procedures for performing systematic reviews.
\newblock {\em Keele, UK, Keele University}, 33(2004):1--26, 2004.

\bibitem{kitchenham2009systematic}
Barbara Kitchenham, O~Pearl Brereton, David Budgen, Mark Turner, John Bailey,
  and Stephen Linkman.
\newblock Systematic literature reviews in software engineering--a systematic
  literature review.
\newblock {\em Information and software technology}, 51(1):7--15, 2009.

\bibitem{replicationpackage}
{Replication package}.
\newblock \url{https://github.com/uOttawa-Nanda-Lab/ML-based-TSP-SLR}.

\bibitem{james2013introduction}
Gareth James, Daniela Witten, Trevor Hastie, and Robert Tibshirani.
\newblock {\em An introduction to statistical learning}, volume 112.
\newblock Springer, 2013.

\bibitem{sutton2018reinforcement}
Richard~S Sutton and Andrew~G Barto.
\newblock {\em Reinforcement learning: An introduction}.
\newblock MIT press, 2018.

\bibitem{manning1999foundations}
Christopher Manning and Hinrich Schutze.
\newblock {\em Foundations of statistical natural language processing}.
\newblock MIT press, 1999.

\bibitem{spieker2017reinforcement}
Helge Spieker, Arnaud Gotlieb, Dusica Marijan, and Morten Mossige.
\newblock Reinforcement learning for automatic test case prioritization and
  selection in continuous integration.
\newblock In {\em Proceedings of the 26th ACM SIGSOFT International Symposium
  on Software Testing and Analysis}, pages 12--22, 2017.

\bibitem{bertolinolearning}
Antonia Bertolino, Antonio Guerriero, Breno Miranda, Roberto Pietrantuono, and
  Stefano Russo.
\newblock Learning-to-rank vs ranking-to-learn: Strategies for regression
  testing in continuous integration.
\newblock In {\em In 42nd International Conference on Software Engineering
  (ICSE)}, 2020.

\bibitem{do2020multi}
Jackson~Antonio do~Prado~Lima and Silvia~Regina Vergilio.
\newblock A multi-armed bandit approach for test case prioritization in
  continuous integration environments.
\newblock {\em IEEE Transactions on Software Engineering}, 2020.

\bibitem{lima2020learning}
Jackson A~Prado Lima, Willian~DF Mendon{\c{c}}a, Silvia~R Vergilio, and
  Wesley~KG Assun{\c{c}}{\~a}o.
\newblock Learning-based prioritization of test cases in continuous integration
  of highly-configurable software.
\newblock In {\em Proceedings of the 24th ACM Conference on Systems and
  Software Product Line: Volume A-Volume A}, pages 1--11, 2020.

\bibitem{lima2020multi}
Jackson A~Prado Lima and Silvia~R Vergilio.
\newblock Multi-armed bandit test case prioritization in continuous integration
  environments: A trade-off analysis.
\newblock In {\em Proceedings of the 5th Brazilian Symposium on Systematic and
  Automated Software Testing}, pages 21--30, 2020.

\bibitem{rosenbauer2020xcs}
Lukas Rosenbauer, Anthony Stein, Roland Maier, David P{\"a}tzel, and J{\"o}rg
  H{\"a}hner.
\newblock Xcs as a reinforcement learning approach to automatic test case
  prioritization.
\newblock In {\em Proceedings of the 2020 Genetic and Evolutionary Computation
  Conference Companion}, pages 1798--1806, 2020.

\bibitem{shi2020reinforcement}
Tingting Shi, Lei Xiao, and Keshou Wu.
\newblock Reinforcement learning based test case prioritization for enhancing
  the security of software.
\newblock In {\em 2020 IEEE 7th International Conference on Data Science and
  Advanced Analytics (DSAA)}, pages 663--672. IEEE, 2020.

\bibitem{almaghairbe2017separating}
Rafig Almaghairbe and Marc Roper.
\newblock Separating passing and failing test executions by clustering
  anomalies.
\newblock {\em Software Quality Journal}, 25(3):803--840, 2017.

\bibitem{carlson2011clustering}
Ryan Carlson, Hyunsook Do, and Anne Denton.
\newblock A clustering approach to improving test case prioritization: An
  industrial case study.
\newblock In {\em ICSM}, volume~11, pages 382--391, 2011.

\bibitem{chen2011using}
Songyu Chen, Zhenyu Chen, Zhihong Zhao, Baowen Xu, and Yang Feng.
\newblock Using semi-supervised clustering to improve regression test selection
  techniques.
\newblock In {\em 2011 Fourth IEEE International Conference on Software
  Testing, Verification and Validation}, pages 1--10. IEEE, 2011.

\bibitem{kandil2017cluster}
Passant Kandil, Sherin Moussa, and Nagwa Badr.
\newblock Cluster-based test cases prioritization and selection technique for
  agile regression testing.
\newblock {\em Journal of Software: Evolution and Process}, 29(6):e1794, 2017.

\bibitem{khalid2019weight}
Zumar Khalid and Usman Qamar.
\newblock Weight and cluster based test case prioritization technique.
\newblock In {\em 2019 IEEE 10th Annual Information Technology, Electronics and
  Mobile Communication Conference (IEMCON)}, pages 1013--1022. IEEE, 2019.

\bibitem{wang2012using}
Yabin Wang, Zhenyu Chen, Yang Feng, Bin Luo, and Yijie Yang.
\newblock Using weighted attributes to improve cluster test selection.
\newblock In {\em 2012 IEEE Sixth International Conference on Software Security
  and Reliability}, pages 138--146. IEEE, 2012.

\bibitem{yoo2009clustering}
Shin Yoo, Mark Harman, Paolo Tonella, and Angelo Susi.
\newblock Clustering test cases to achieve effective and scalable
  prioritisation incorporating expert knowledge.
\newblock In {\em Proceedings of the eighteenth international symposium on
  Software testing and analysis}, pages 201--212, 2009.

\bibitem{busjaeger2016learning}
Benjamin Busjaeger and Tao Xie.
\newblock Learning for test prioritization: an industrial case study.
\newblock In {\em Proceedings of the 2016 24th ACM SIGSOFT International
  Symposium on Foundations of Software Engineering}, pages 975--980, 2016.

\bibitem{chen2018optimizing}
Junjie Chen, Yiling Lou, Lingming Zhang, Jianyi Zhou, Xiaoleng Wang, Dan Hao,
  and Lu~Zhang.
\newblock Optimizing test prioritization via test distribution analysis.
\newblock In {\em Proceedings of the 2018 26th ACM Joint Meeting on European
  Software Engineering Conference and Symposium on the Foundations of Software
  Engineering}, pages 656--667, 2018.

\bibitem{hasnain2019recurrent}
Muhammad Hasnain, Muhammad~Fermi Pasha, Chern~Hong Lim, and Imran Ghan.
\newblock Recurrent neural network for web services performance forecasting,
  ranking and regression testing.
\newblock In {\em 2019 Asia-Pacific Signal and Information Processing
  Association Annual Summit and Conference (APSIPA ASC)}, pages 96--105. IEEE,
  2019.

\bibitem{jahan2019version}
Hosney Jahan, Ziliang Feng, SM~Mahmud, and Penglin Dong.
\newblock Version specific test case prioritization approach based on
  artificial neural network.
\newblock {\em Journal of Intelligent \& Fuzzy Systems}, 36(6):6181--6194,
  2019.

\bibitem{lachmann2016system}
Remo Lachmann, Sandro Schulze, Manuel Nieke, Christoph Seidl, and Ina Schaefer.
\newblock System-level test case prioritization using machine learning.
\newblock In {\em 2016 15th IEEE International Conference on Machine Learning
  and Applications (ICMLA)}, pages 361--368. IEEE, 2016.

\bibitem{mahdieh2020incorporating}
Mostafa Mahdieh, Seyed-Hassan Mirian-Hosseinabadi, Khashayar Etemadi, Ali
  Nosrati, and Sajad Jalali.
\newblock Incorporating fault-proneness estimations into coverage-based test
  case prioritization methods.
\newblock {\em Information and Software Technology}, 121:106269, 2020.

\bibitem{mirarab2008empirical}
Siavash Mirarab and Ladan Tahvildari.
\newblock An empirical study on bayesian network-based approach for test case
  prioritization.
\newblock In {\em 2008 1st International Conference on Software Testing,
  Verification, and Validation}, pages 278--287. IEEE, 2008.

\bibitem{noor2017studying}
Tanzeem~Bin Noor and Hadi Hemmati.
\newblock Studying test case failure prediction for test case prioritization.
\newblock In {\em Proceedings of the 13th International Conference on
  Predictive Models and Data Analytics in Software Engineering}, pages 2--11,
  2017.

\bibitem{palma2018improvement}
Francis Palma, Tamer Abdou, Ayse Bener, John Maidens, and Stella Liu.
\newblock An improvement to test case failure prediction in the context of test
  case prioritization.
\newblock In {\em Proceedings of the 14th International Conference on
  Predictive Models and Data Analytics in Software Engineering}, pages 80--89,
  2018.

\bibitem{article}
Mayank~Mohan Sharma and Akshat Agrawal.
\newblock Test case design and test case prioritization using machine learning.
\newblock {\em International Journal of Engineering and Advanced Technology},
  9(1):2742--2748, 2019.

\bibitem{singhmachine}
Ajmer Singh, Rajesh~Kumar Bhatia, and Anita Singhrova.
\newblock Machine learning based test case prioritization in object oriented
  testing.
\newblock {\em International Journal of Recent Technology and Engineering},
  8(3):700--707, 2019.

\bibitem{tonella2006using}
Paolo Tonella, Paolo Avesani, and Angelo Susi.
\newblock Using the case-based ranking methodology for test case
  prioritization.
\newblock In {\em 2006 22nd IEEE International Conference on Software
  Maintenance}, pages 123--133, 2006.

\bibitem{aman2020comparative}
Hirohisa Aman, Sousuke Amasaki, Tomoyuki Yokogawa, and Minoru Kawahara.
\newblock A comparative study of vectorization-based static test case
  prioritization methods.
\newblock In {\em 2020 46th Euromicro Conference on Software Engineering and
  Advanced Applications (SEAA)}, pages 80--88. IEEE, 2020.

\bibitem{medhat2020framework}
Noha Medhat, Sherin~M Moussa, Nagwa~Lotfy Badr, and Mohamed~F Tolba.
\newblock A framework for continuous regression and integration testing in iot
  systems based on deep learning and search-based techniques.
\newblock {\em IEEE Access}, 8:215716--215726, 2020.

\bibitem{thomas2014static}
Stephen~W Thomas, Hadi Hemmati, Ahmed~E Hassan, and Dorothea Blostein.
\newblock Static test case prioritization using topic models.
\newblock {\em Empirical Software Engineering}, 19(1):182--212, 2014.

\bibitem{wilson1995classifier}
Stewart~W Wilson.
\newblock Classifier fitness based on accuracy.
\newblock {\em Evolutionary computation}, 3(2):149--175, 1995.

\bibitem{li2011learning}
Hang Li.
\newblock Learning to rank for information retrieval and natural language
  processing.
\newblock {\em Synthesis Lectures on Human Language Technologies}, 4(1):1--113,
  2011.

\bibitem{Ranklibs}
Van Dang and Michael Zarozinski.
\newblock Ranklib.
\newblock \url{https://sourceforge.net/p/lemur/wiki/RankLib/}, 2020.

\bibitem{SVMRank}
Thorsten Joachims.
\newblock Optimizing search engines using clickthrough data.
\newblock In {\em Proceedings of the eighth ACM SIGKDD international conference
  on Knowledge discovery and data mining}, pages 133--142, 2002.

\bibitem{freund2003efficient}
Yoav Freund, Raj Iyer, Robert~E Schapire, and Yoram Singer.
\newblock An efficient boosting algorithm for combining preferences.
\newblock {\em Journal of machine learning research}, 4(Nov):933--969, 2003.

\bibitem{yue2007support}
Yisong Yue, Thomas Finley, Filip Radlinski, and Thorsten Joachims.
\newblock A support vector method for optimizing average precision.
\newblock In {\em Proceedings of the 30th annual international ACM SIGIR
  conference on Research and development in information retrieval}, pages
  271--278, 2007.

\bibitem{zhang2019incremental}
Chongsheng Zhang, Yuan Zhang, Xianjin Shi, George Almpanidis, Gaojuan Fan, and
  Xiajiong Shen.
\newblock On incremental learning for gradient boosting decision trees.
\newblock {\em Neural Processing Letters}, 50(1):957--987, 2019.

\bibitem{LSTM}
Sepp Hochreiter and J{\"u}rgen Schmidhuber.
\newblock Long short-term memory.
\newblock {\em Neural computation}, 9(8):1735--1780, 1997.

\bibitem{alon2019code2vec}
Uri Alon, Meital Zilberstein, Omer Levy, and Eran Yahav.
\newblock code2vec: Learning distributed representations of code.
\newblock {\em Proceedings of the ACM on Programming Languages}, 3(POPL):1--29,
  2019.

\bibitem{nunez2017source}
Alberto~S Nu{\~n}ez-Varela, H{\'e}ctor~G P{\'e}rez-Gonzalez, Francisco~E
  Mart{\'\i}nez-Perez, and Carlos Soubervielle-Montalvo.
\newblock Source code metrics: A systematic mapping study.
\newblock {\em Journal of Systems and Software}, 128:164--197, 2017.

\bibitem{JaCoCo}
{EclEmma team}.
\newblock {JaCoCo: Java Code Coverage Library}.
\newblock \url{https://github.com/jacoco/jacoco}, 2021.
\newblock Retrieved March 14, 2021.

\bibitem{Clover}
{Atlassian Clover}.
\newblock {Atlassian Clover}.
\newblock \url{https://bitbucket.org/atlassian/clover}, 2021.
\newblock Retrieved March 14, 2021.

\bibitem{GCOV}
{Nicholas McGuire}.
\newblock {Linux Kernel GCOV - tool analysis}.
\newblock \url{https://linuxdevices.org/ldfiles/article062/der_herr_gcov.pdf},
  2006.
\newblock Retrieved March 14, 2021.

\bibitem{msr17challenge}
Moritz Beller, Georgios Gousios, and Andy Zaidman.
\newblock Travistorrent: Synthesizing travis ci and github for full-stack
  research on continuous integration.
\newblock In {\em 2017 IEEE/ACM 14th International Conference on Mining
  Software Repositories (MSR)}, pages 447--450. IEEE, 2017.

\bibitem{elbaum2002test}
Sebastian Elbaum, Alexey~G Malishevsky, and Gregg Rothermel.
\newblock Test case prioritization: A family of empirical studies.
\newblock {\em IEEE transactions on software engineering}, 28(2):159--182,
  2002.

\bibitem{qu2007combinatorial}
Xiao Qu, Myra~B Cohen, and Katherine~M Woolf.
\newblock Combinatorial interaction regression testing: A study of test case
  generation and prioritization.
\newblock In {\em 2007 IEEE International Conference on Software Maintenance},
  pages 255--264. IEEE, 2007.

\bibitem{elbaum2001incorporating}
Sebastian Elbaum, Alexey Malishevsky, and Gregg Rothermel.
\newblock Incorporating varying test costs and fault severities into test case
  prioritization.
\newblock In {\em Proceedings of the 23rd International Conference on Software
  Engineering. ICSE 2001}, pages 329--338. IEEE, 2001.

\bibitem{just2014defects4j}
Ren{\'e} Just, Darioush Jalali, and Michael~D Ernst.
\newblock Defects4j: A database of existing faults to enable controlled testing
  studies for java programs.
\newblock In {\em Proceedings of the 2014 International Symposium on Software
  Testing and Analysis}, pages 437--440, 2014.

\bibitem{do2006use}
Hyunsook Do and Gregg Rothermel.
\newblock On the use of mutation faults in empirical assessments of test case
  prioritization techniques.
\newblock {\em IEEE Transactions on Software Engineering}, 32(9):733--752,
  2006.

\bibitem{luo2018assessing}
Qi~Luo, Kevin Moran, Denys Poshyvanyk, and Massimiliano Di~Penta.
\newblock Assessing test case prioritization on real faults and mutants.
\newblock In {\em 2018 IEEE international conference on software maintenance
  and evolution (ICSME)}, pages 240--251. IEEE, 2018.

\bibitem{yan2010dynamic}
Shali Yan, Zhenyu Chen, Zhihong Zhao, Chen Zhang, and Yuming Zhou.
\newblock A dynamic test cluster sampling strategy by leveraging execution
  spectra information.
\newblock In {\em 2010 Third International Conference on Software Testing,
  Verification and Validation}, pages 147--154. IEEE, 2010.

\bibitem{robbins1952some}
Herbert Robbins.
\newblock Some aspects of the sequential design of experiments.
\newblock {\em Bulletin of the American Mathematical Society}, 58(5):527--535,
  1952.

\bibitem{wilks2011statistical}
Daniel~S Wilks.
\newblock {\em {Statistical methods in the atmospheric sciences}}, volume 100.
\newblock Academic press, 2011.

\bibitem{cliff1993dominance}
Norman Cliff.
\newblock {Dominance statistics: Ordinal analyses to answer ordinal questions}.
\newblock {\em Psychological Bulletin}, 114(3):494, 1993.

\bibitem{stone1974cross}
Mervyn Stone.
\newblock Cross-validatory choice and assessment of statistical predictions.
\newblock {\em Journal of the Royal Statistical Society: Series B
  (Methodological)}, 36(2):111--133, 1974.

\bibitem{liem2020run}
Cynthia Liem and Annibale Panichella.
\newblock Run, forest, run? on randomization and reproducibility in predictive
  software engineering.
\newblock {\em arXiv preprint arXiv:2012.08387}, 2020.

\bibitem{mattis2020rtptorrent}
Toni Mattis, Patrick Rein, Falco D{\"u}rsch, and Robert Hirschfeld.
\newblock Rtptorrent: An open-source dataset for evaluating regression test
  prioritization.
\newblock In {\em Proceedings of the 17th International Conference on Mining
  Software Repositories}, pages 385--396, 2020.

\bibitem{devlin2018bert}
Jacob Devlin, Ming-Wei Chang, Kenton Lee, and Kristina Toutanova.
\newblock {BERT: Pre-training of deep bidirectional transformers for language
  understanding}.
\newblock {\em arXiv preprint arXiv:1810.04805}, 2018.

\bibitem{feng2020codebert}
Zhangyin Feng, Daya Guo, Duyu Tang, Nan Duan, Xiaocheng Feng, Ming Gong, Linjun
  Shou, Bing Qin, Ting Liu, Daxin Jiang, et~al.
\newblock {CodeBERT: A pre-trained model for programming and natural
  languages}.
\newblock {\em arXiv preprint arXiv:2002.08155}, 2020.

\bibitem{kazmi2017effective}
Rafaqut Kazmi, Dayang~NA Jawawi, Radziah Mohamad, and Imran Ghani.
\newblock Effective regression test case selection: A systematic literature
  review.
\newblock {\em ACM Computing Surveys (CSUR)}, 50(2):1--32, 2017.

\bibitem{durelli2019machine}
Vinicius~HS Durelli, Rafael~S Durelli, Simone~S Borges, Andre~T Endo, Marcelo~M
  Eler, Diego~RC Dias, and Marcelo~P Guimaraes.
\newblock Machine learning applied to software testing: A systematic mapping
  study.
\newblock {\em IEEE Transactions on Reliability}, 68(3):1189--1212, 2019.

\bibitem{shaheamlung2020comprehensive}
Golmei Shaheamlung, Ketusezo Rote, et~al.
\newblock A comprehensive review for test case prioritization in software
  engineering.
\newblock In {\em 2020 International Conference on Intelligent Engineering and
  Management (ICIEM)}, pages 331--336. IEEE, 2020.

\end{thebibliography}

\newpage
\section*{Authors' Biographies}
\begingroup
\setlength{\intextsep}{-1.5pt}
\begin{wrapfigure}{l}{32.5mm} 
    \includegraphics[width=1.8in,height=1.5in,clip,keepaspectratio]{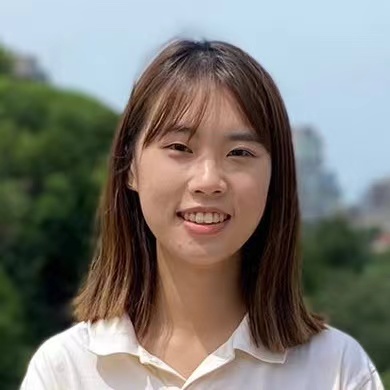}
  \end{wrapfigure}\par
  \textbf{Rongqi Pan} is a second-year PhD student at the School of EECS at the University of Ottawa and a member of Nanda Lab. She obtained her Master’s Degree in Statistics at the University of Illinois at Urbana-Champaign. Her research interests centre around machine learning and software testing. Her current research is focused on optimizing software regression testing in the context of continuous integration.\par
\endgroup

\vspace{52pt}

\begingroup
\setlength{\intextsep}{-1.5pt}
\begin{wrapfigure}{l}{32.5mm} 
    \includegraphics[width=2.3in,height=1.73in,clip,keepaspectratio]{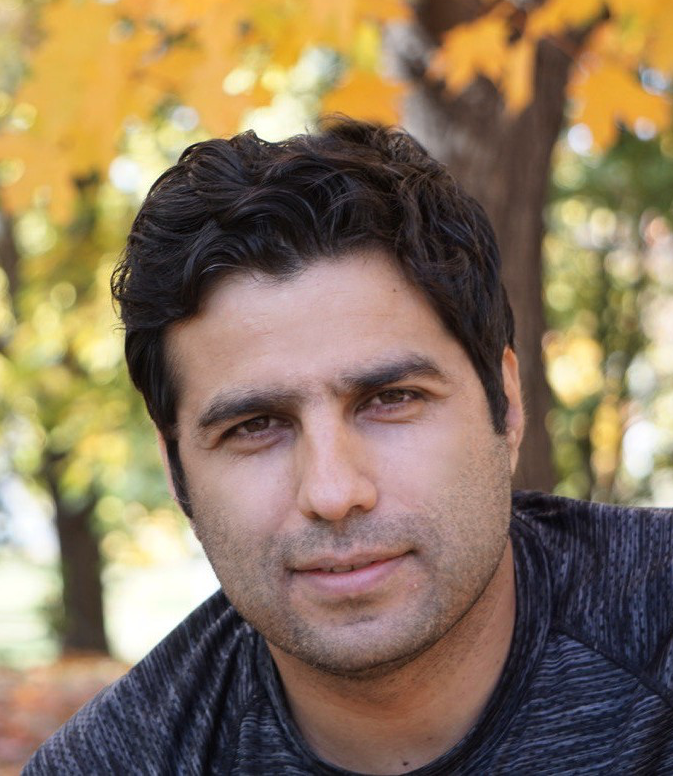}
  \end{wrapfigure}\par
  \noindent\textbf{Mojtaba Bagherzadeh} is a Postdoctoral Research Fellow at the School of EECS at the University of Ottawa. He obtained his Ph.D. in Computer Science from Queen’s University, Canada (2019). He has several years of experience working as a software developer at IBM and a startup company. His research interests include testing and debugging of machine learning-based systems, model driven engineering, software testing, and empirical software engineering.\par
\endgroup

\vspace{57pt}

\begingroup
\setlength{\intextsep}{-1.5pt}
\begin{wrapfigure}{l}{32.5mm} 
    \includegraphics[width=2.3in,height=1.73in,clip,keepaspectratio]{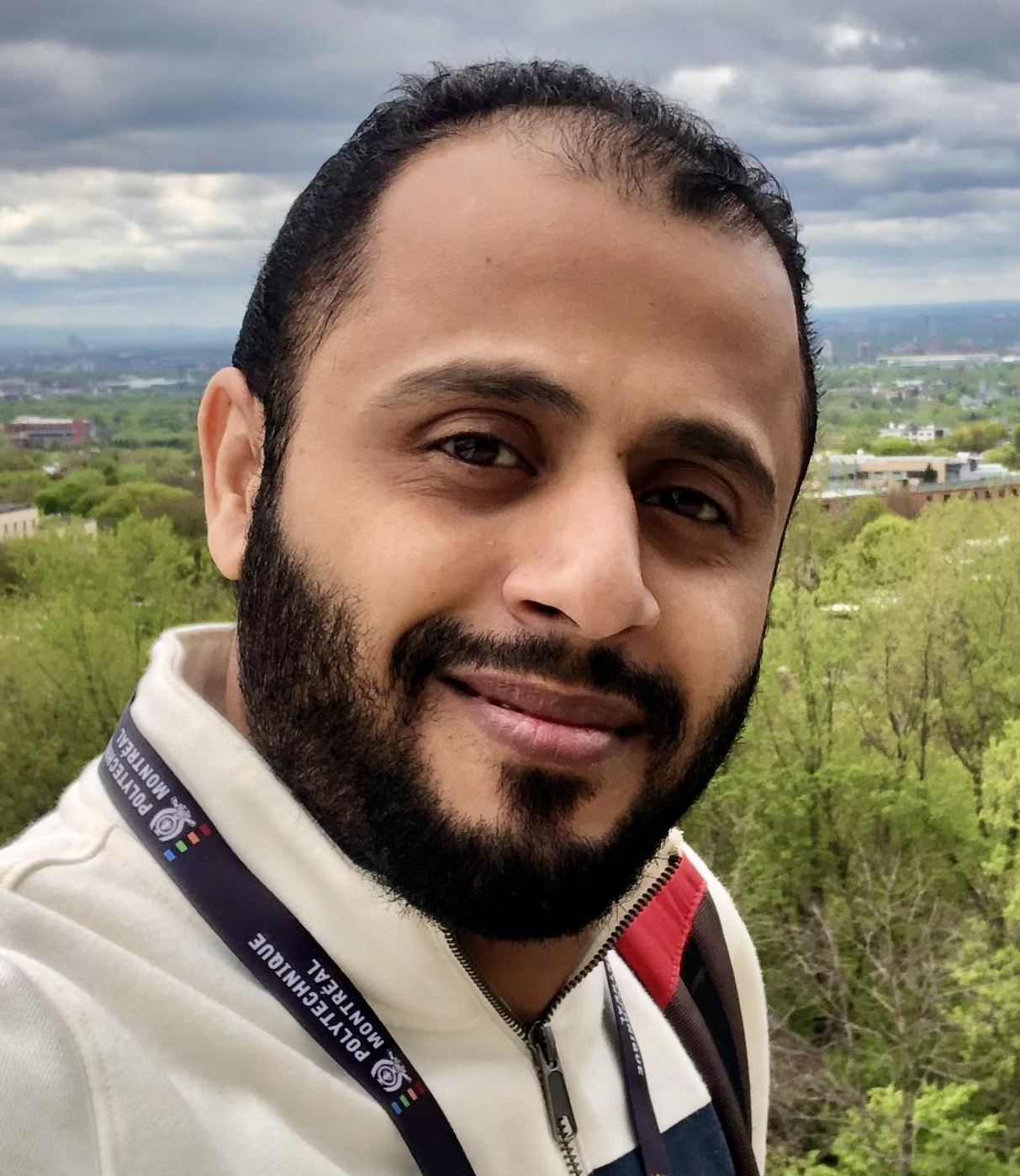}
  \end{wrapfigure}\par
  \noindent\textbf{Taher A. Ghaleb} is a Postdoctoral Research Fellow at the School of EECS at the University of Ottawa, Canada. He obtained his Ph.D. in Computing from Queen’s University, Canada, in 2021. During his Ph.D., Taher held an Ontario Trillium Scholarship, a highly prestigious award for doctoral students. He has been working as a research/teaching assistant since he obtained his B.Sc. in Information Technology from Taiz University, Yemen (2008) and M.Sc. in Computer Science from King Fahd University of Petroleum and Minerals, Saudi Arabia (2016). His research interests include continuous integration, software testing, mining software repositories, applied machine learning, program analysis, and empirical software engineering.\par
\endgroup

\vspace{12pt}

\begingroup
\setlength{\intextsep}{-1.5pt}
\begin{wrapfigure}{l}{32.5mm} 
    \includegraphics[width=2in,height=1.5in,clip,keepaspectratio]{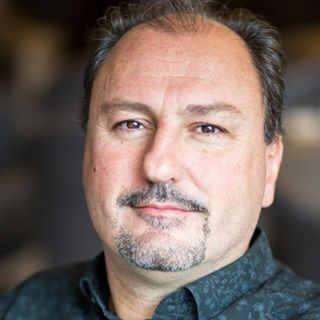}
  \end{wrapfigure}\par
  \noindent\textbf{Lionel C. Briand} is professor of software engineering and has shared appointments between (1) School of Electrical Engineering and Computer Science, University of Ottawa, Canada and (2) The SnT centre for Security, Reliability, and Trust, University of Luxembourg. He is the head of the SVV department at the SnT Centre and a Canada Research Chair in Intelligent Software Dependability and Compliance (Tier 1). He holds an ERC Advanced Grant, the most prestigious European individual research award, and has conducted applied research in collaboration with industry for more than 25 years, including projects in the automotive, aerospace, manufacturing, financial, and energy domains. He was elevated to the grades of IEEE and ACM fellow, granted the IEEE Computer Society Harlan Mills award (2012) and the IEEE Reliability Society Engineer-of-the-year award (2013) for his work on software verification and testing. His research interests include: Testing and verification, search-based software engineering, model-driven development, requirements engineering, and empirical software engineering.\par
\endgroup

\end{document}